\documentclass[useAMS,usenatbib]{mn2e}
\usepackage{amssymb}
\usepackage{deluxetable}
\usepackage{url}
\usepackage{lineno}
\newcommand{\G}{\Gamma}

\newcommand{\e}{\epsilon}

\newcommand{\cm}{\mathrm{cm}}

\newcommand{\erg}{\mathrm{erg}}

\newcommand{\s}{\mathrm{s}}

\title[Powerful jets in NLSy1s]{Investigating powerful jets in radio-loud Narrow Line Seyfert 1s} 
\author[M. Orienti, F. D'Ammando, J. Larsson et al.]
  {M. Orienti$^{1}$\thanks{E-mail: orienti@ira.inaf.it},
F. D'Ammando$^{1,2}$, J. Larsson$^{3}$, J. Finke$^{4}$, M. Giroletti$^{1}$,
\newauthor D. Dallacasa$^{1,2}$,
T. Isacsson$^{3}$, J. Stoby Hoglund$^{3}$\\
$^1$INAF -- Istituto di Radioastronomia, via Gobetti 101, I-40129, Bologna,
Italy \\
$^{2}$Dipartimento di Fisica e Astronomia, Universi\`a degli Studi di
Bologna, via Ranzani 1, I-40127 Bologna, Italy\\
$^{3}$KTH, Department of Physics, and the Oskar Klein Centre, AlbaNova, SE-106 91 Stockholm, Sweden\\ 
$^{4}$ US Naval Research Laboratory, Code 7653, 4555 Overlook Avenue SW, Washington, DC 20375-5352, USA\\
}
\date{Received \today; accepted ?}

\pagerange{\pageref{firstpage}--\pageref{lastpage}} \pubyear{2002}

\def\LaTeX{L\kern-.36em\raise.3ex\hbox{a}\kern-.15em
    T\kern-.1667em\lower.7ex\hbox{E}\kern-.125emX}

\begin{document}

\label{firstpage}

\maketitle

\begin{abstract}
We report results on multiband observations from radio to
$\gamma$-rays of the two radio-loud narrow-line
Seyfert~1 (NLSy1) galaxies PKS\,2004$-$447 and J1548$+$3511. Both sources show a
core-jet structure on parsec  
scale, while they are unresolved at the arcsecond scale. 
The high
core dominance and the high variability
brightness temperature make these NLSy1 
galaxies good $\gamma$-ray source candidates. 
{\it Fermi}-LAT detected $\gamma$-ray emission only from
PKS\,2004$-$447, with a $\gamma$-ray luminosity comparable to that
observed in blazars. No $\gamma$-ray emission is observed for
J1548$+$3511. Both sources are variable in X-rays. J1548$+$3511
shows a hardening 
of the spectrum during high activity states, while PKS\,2004$-$447 has
no spectral variability. A spectral steepening likely related to
  the soft excess is hinted
below 2 keV for J1548$+$3511, while the X-ray spectra of
PKS\,2004$-$447 collected by {\it XMM-Newton} in 2012 are described by
a single power-law without significant soft excess.
No additional absorption above the
Galactic column density or the presence of an Fe line is detected in the
X-ray spectra of both sources. 

\end{abstract}


\section{Introduction}

Narrow Line Seyfert 1 (NLSy1) galaxies represent a rare type of
classical Seyfert galaxies. The strong featureless X-ray continuum and the
strong high ionization lines shown by NLSy1 are common in Seyfert
1. However, the  
optical permitted emission lines are narrow, i.e. more similar to
Seyfert 2 galaxies, indicating a combination of properties from both types.
Their optical spectra are characterized by narrow permitted
lines (full width at half-maximum, FWHM
$\leq$ 2000 km s$^{-1}$), weak [O$_{\rm III}$]/$\lambda$5007 emission line
([O$_{\rm III}$]/H$\beta < 3$) and usually strong Fe$_{\rm II}$ emission lines
\citep{osterbrock85}. \\
NLSy1 are usually hosted in spiral galaxies, although some objects are
associated with early-type S0 galaxies (see e.g. Mrk\,705 and
Mrk\,1239; Markarian et al. 1989). \\
Extreme characteristics are observed in the X-ray band, where strong
and rapid variability is observed more frequently than in classical
Seyfert~1. The X-ray spectrum of NLSy1 is usually described by a
power law 
which dominates in 
the 2--10 keV energy range, and a soft X-ray excess at lower energies.
The X-ray spectrum between 0.3 and 10 keV is steeper than in Seyfert~1
\citep{grupe10}, 
while the photon index of the hard X-ray spectrum is similar 
\citep{panessa11}. 
The complex X-ray spectrum is 
interpreted in terms of either relativistic blurred disc reflection,
or ionized/neutral absorption covering the X-ray source
\citep{fabian09,miller10}. \\
About 7 per cent of NLSy1 are radio-loud (RL), with a
smaller fraction ($\sim$2.5 per cent) exhibiting a high radio-loudness parameter\footnote{The radio-loudness is defined as the ratio between the
radio flux density at 6 cm and the optical B-band flux \citep{kellermann89}.}
(${\rm R} >100$; Komossa et al. 2006).
In the radio band, RL-NLSy1 usually show a compact morphology with a
one-sided jet emerging from the bright core and extending up to a few
parsecs. In some objects the radio emission extends on kpc scales
\citep{richards15,doi12,anton08}. The
high values of both brightness temperature and core 
dominance suggest the presence of non-thermal emission from
relativistic jets \citep{doi11}. The measurement of superluminal
motion in the RL-NLSy1 SBS\,0846$+$513 indicates Doppler boosting
effects in relativistic jets \citep{dammando13b}.
It is worth noting that some evidence of pc-scale jet-like
structure is also found in some radio-quiet (RQ) NLSy1, but the nature of the
outflow is still under debate \citep[e.g.][]{doi13}. \\
Strong evidence in favour of highly-relativistic jets in RL-NLSy1 is
the detection by the Large Area Telescope (LAT) on board the {\it
  Fermi} satellite of $\gamma$-ray emission from 6 RL-NLSy1: 
PMN\,J0948$+$0022, 1H\,0323$+$342, PKS\,1502$+$036, PKS\,2004$-$447
\citep{abdo09}, SBS\,0846$+$513 \citep{dammando12}, and
FBQS\,J1644$+$2619 \citep{dammando15b}. The $\gamma$-ray emission is
variable, showing flaring activity accompanied by a moderate hardening of the
spectrum. The peculiar multiwavelength properties together with the
$\gamma$-ray flares make the RL-NLSy1 more similar to blazars rather
than classical Seyfert galaxies, at least at high energies.\\
Relativistic jets produced by nuclear objects which are thought to be
hosted mainly in spiral 
galaxies are 
somehow puzzling. RL-NLSy1 are usually found at higher redshift than
RQ-NLSy1 and no optical morphological studies have been carried out so
far, with the exception of 1H\,0323$+$342. The optical morphology of
the host galaxy is compatible with either a one-armed spiral
\citep{zhou07} or a
ring-like structure produced by a recent merger
\citep{leon14,anton08}.\\
In this paper we present results of a multiwavelength study, from
radio to $\gamma$-rays, of the
RL-NLSy1 J1548$+$3511 and PKS\,2004$-$447. 
J1548$+$3511 is a NLSy1 at redshift $z=0.478$ with a radio loudness
R$\sim$110 and an 
estimated black hole mass $M_{\rm BH} = 10^{7.9} M_{\odot}$
\citep{yuan08}, while PKS\,2004$-$447, at redshift $z=0.24$, has
$1710< {\rm R} < 6320$ and the
estimated black hole mass is 10$^{6.7} M_{\odot}$ \citep{oshlack01}. 
These sources have been selected on the basis of their high
  variability brightness temperature T'$_{\rm B, var} = 10^{13}$ -- $10^{15}
  K$ that is considered a good indicator for the presence of Doppler boosting in
  relativistic jets. Among the RL-NLSy1
  presented in the sample by \citet{yuan08}, J1548$+$3511 is the only
  source with  T'$_{\rm B,
    var}$ as high as those found in $\gamma$-ray-loud NLSy1. However,
  no $\gamma$-ray emission has been detected from this source so
  far. In this paper we aim at investigating differences and
  similarities between the $\gamma$-ray emitter PKS\,2004$-$447 and
  the $\gamma$-ray silent J1548$+$3511.  
The results for these two sources are then
compared to what has been found for the other $\gamma$-ray emitting 
RL-NLSy1 in the
literature, in order to investigate the peculiarity of this sub-class
of objects. The information from the
  multiwavelength data of J1548$+$3511 and PKS\,2004$-447$ is then
  used to model the spectral energy 
  distribution (SED) of these two sources.\\
The paper is organized as follows: in Sections 2, 3, 4 and 5 we report
the radio, $\gamma$-ray {\it Fermi}-LAT, X-ray {\it Swift} and {\it
  XMM-Newton} data analysis. In Section 6 we present the results from
the multiwavelength analysis. The discussion and the presentation of
the SED modelling are given in Section 7, while a brief summary is in
Section 8.\\  
Throughout this paper, we assume the following cosmology: $H_{0} =
71\; {\rm km \; s^{-1} \; Mpc^{-1}}$, $\Omega_{\rm M} = 0.27$ and
$\Omega_{\rm \Lambda} = 0.73$, in a flat Universe. \\
%

\section{Radio data}

\subsection{VLBA observations}
\label{vlba_obs}

\begin{figure*}
\begin{center}
\includegraphics{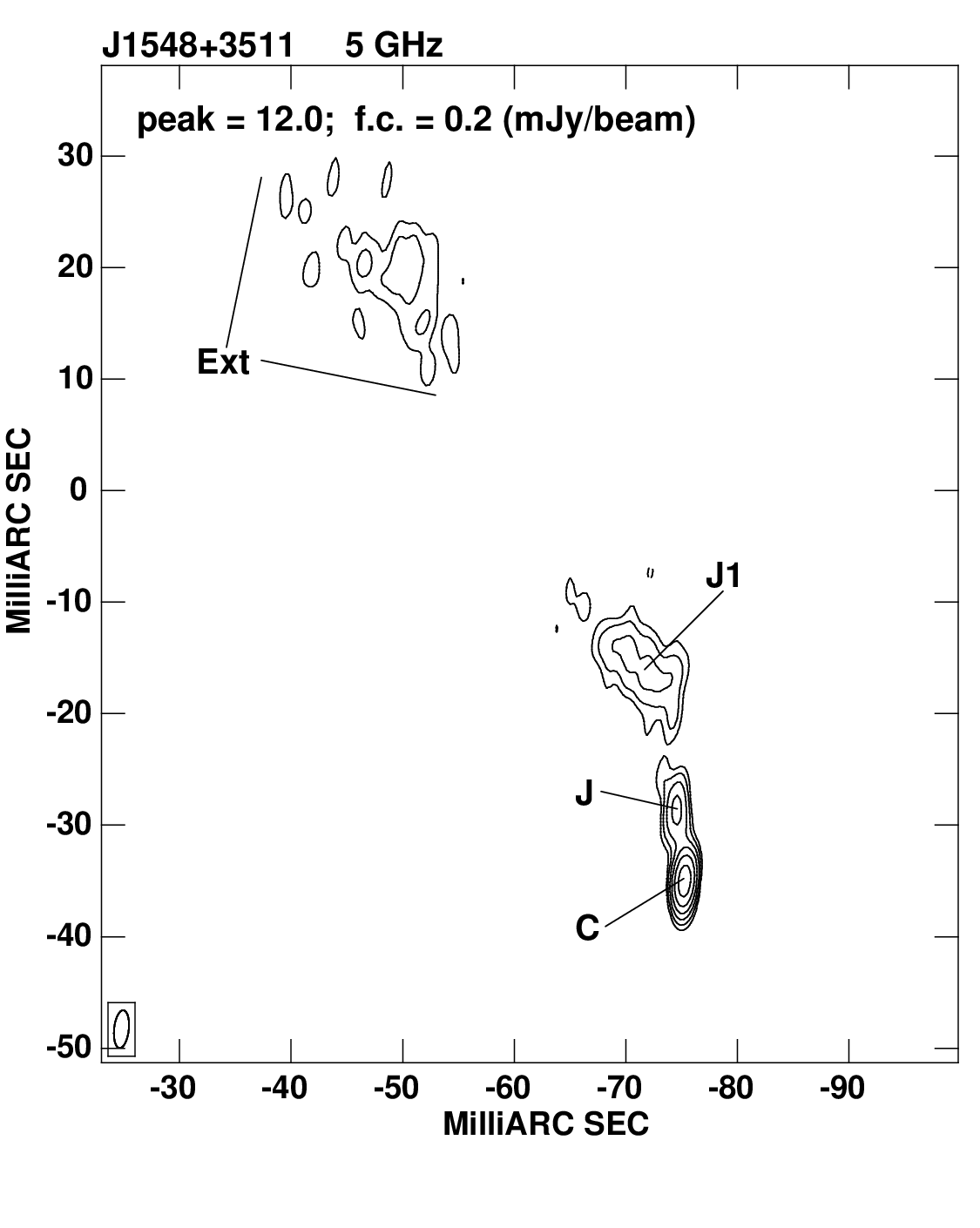}
\includegraphics{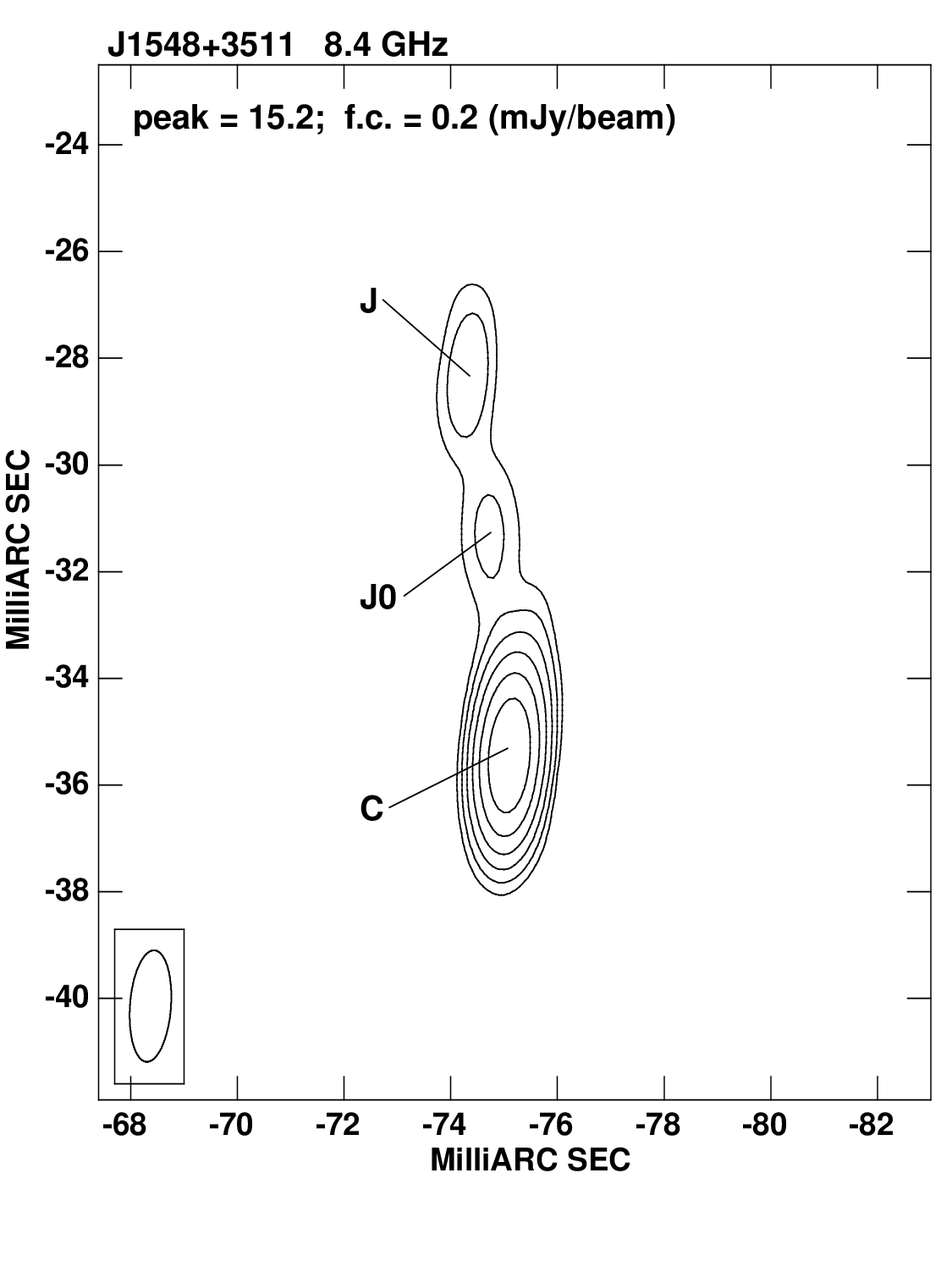}
\includegraphics{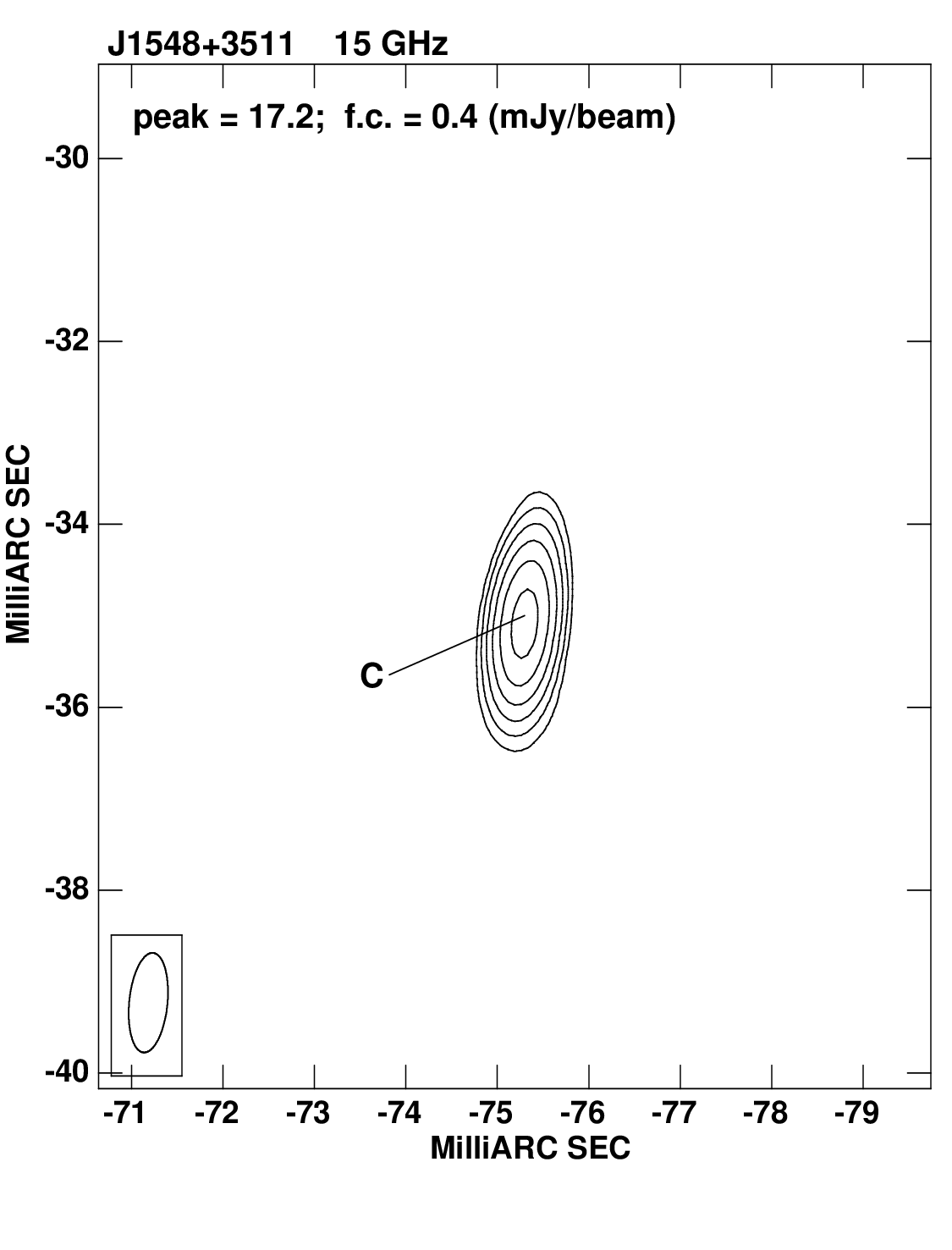}
\vspace{8cm}
\caption{VLBA images of J1548$+$3511 at 5 GHz ({\it left}), 8.4 GHz
  ({\it centre}), and 15 GHz ({\it right}). On each image we provide 
the peak flux density in mJy beam$^{-1}$ and the first contour
intensity (f.c.) in mJy beam$^{-1}$, which corresponds to three times
the off-source noise level measured on the image plane. 
Contours increase by a factor of 2. The restoring beam 
is plotted in the bottom left corner of each panel. }
\label{1548_vlba}
\end{center}
\end{figure*}

\begin{figure}
\begin{center}
\includegraphics{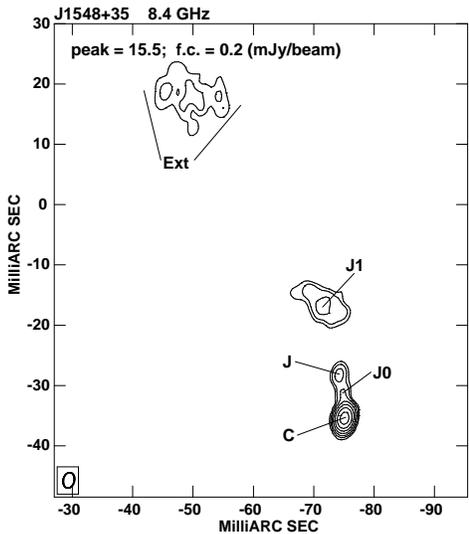}
\vspace{7cm}
\caption{Low-resolution VLBA image of J1548$+$3511 at 
8.4 GHz. On the image we provide 
the peak flux density in mJy beam$^{-1}$ and the first contour
intensity (f.c.) in mJy beam$^{-1}$, which corresponds to two times
the off-source noise level measured on the image plane. 
Contours increase by a factor of 2. The restoring beam 
is plotted in the bottom left corner. }
\label{1548_natural}
\end{center}
\end{figure}

\begin{figure}
\begin{center}
\includegraphics{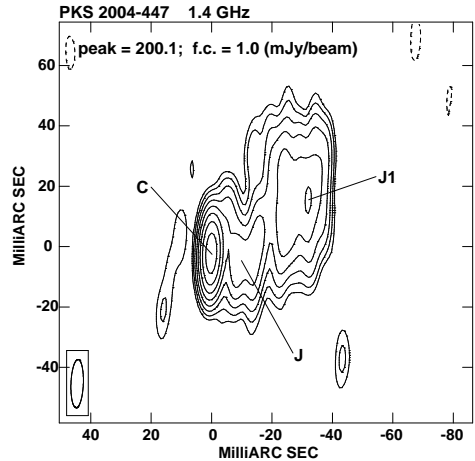}
\vspace{6.5cm}
\caption{VLBA image at 1.4 GHz of PKS\,2004$-$447. On the image we provide 
the peak flux density in mJy beam$^{-1}$ and the first contour
intensity (f.c.) in mJy beam$^{-1}$, which corresponds to three times
the off-source noise level measured on the image plane. 
Contours increase by a factor of 2. The restoring beam 
is plotted in the bottom left corner. }
\label{2004_vlba}
\end{center}
\end{figure}

Multifrequency Very Long Baseline Array (VLBA) observation (project code
BO045) of J1548$+$3511 was carried out at 
5, 8.4, and 15 GHz on 2013, January 2. The observation was performed
in phase reference mode
with a recording bandwidth of 16 MHz per channel in dual
polarization at 512 
Mbps data rate. 
The target source was observed for 45 min at 5
and 8.4 GHz, and for 75 min at 15 GHz, spread into several scans and cycling 
through frequencies and calibrators in order to improve the {\it  uv}-coverage. The strong source 3C\,345 was used as fringe finder
and bandpass calibrator. The source J1602$+$3326 was used as phase
calibrator. \\ 
Data reduction was performed using the NRAO's Astronomical Image
Processing System (\texttt{AIPS}). A priori amplitude calibration was
derived using measurements of the system temperature and the antenna
gains. 
The uncertainties on the amplitude calibration ($\sigma_{c}$) were found 
to be approximately 
5 per cent at 5 and 8.4 GHz, and about 7 per cent at 15 GHz.
Final images were produced
after a number of phase self-calibration iterations (Fig. \ref{1548_vlba}). 
The 1$\sigma$ noise (rms) level measured on the 
image plane is about 0.08 mJy beam$^{-1}$ at 5 and 8.4 GHz,
and about 0.12 
mJy beam$^{-1}$ at 15 GHz. The restoring beam is 3.4$\times$1.3 mas$^{2}$,
2.1$\times$0.8 mas$^{2}$, and 1.1$\times$0.4 mas$^{2}$ at 5, 8.4 and
15 GHz, respectively.
In addition to the full-resolution images, at 5 and 8.4 GHz we
produced ``low-resolution'' images using natural grid weighting and a
maximum baseline of 100 M$\lambda$. The low-resolution 
image at 8.4 GHz is presented in Fig. \ref{1548_natural}. The
restoring beam is 2.6$\times$1.8 mas$^{2}$.\\ 
To study the parsec-scale structure of PKS\,2004$-$447 we retrieved
archival VLBA data at 1.4 GHz (project code BD050). 
The observation was performed on 1998, October 13 with a recording
bandwidth of 8 MHz per channel in dual polarization at 256 Mbps
data rate. In this observing run 
PKS\,2004$-$447 was observed as phase calibrator, for a total time of
about 30 minutes. Seven VLBA antennas
participated in the observing run. Due to the southern
declination of 
the source the restoring beam is highly elongated in the North-South
direction and is
16.0$\times$4.1 mas$^{2}$.\\
The data were reduced following the same procedure described for
J1548$+$3511. 
%
%
The uncertainties on the amplitude calibration are $\sigma_{c}$ $\sim$
10 per cent. 
Final images were produced
after a number of phase self-calibration iterations. Amplitude
self-calibration was applied using a solution interval longer than the
scan length to remove residual systematic errors at the end of the
self-calibration process. The final image is presented in
Fig. \ref{2004_vlba}. The 1$\sigma$ noise level measured on the
image plane is $\sim$0.3 mJy beam$^{-1}$. \\
The total flux density of each source was measured by using the
\texttt{AIPS} verb TVSTAT, which performs an aperture integration over
a selected region on the image plane. In case of bright and compact components,
like the core, we used the task JMFIT, which performs a Gaussian fit
to the source component on the image plane. For more extended
sub-components, like jets, the flux density was measured by TVSTAT.
The uncertainty in the flux density arises from both the calibration
error $\sigma_{c}$ and the measurement error
$\sigma_{m}$. The latter represents the off-source rms noise level
measured on the image plane and is related to the source size
$\theta_{\rm obs}$ normalized by the beam size $\theta_{\rm beam}$ as 
$\sigma_{m} = {\rm rms} \times$($\theta_{\rm obs}/ \theta_{\rm beam}$)$^{1/2}$.
The flux density error $\sigma_{\rm S}$ reported in Table
\ref{flux_vlba} takes both uncertainties into account and corresponds
to $\sigma_{\rm S} = \sqrt{\sigma_{c}^{2} + \sigma_{m}^{2}}$. \\

\subsection{Archival VLA data}

To investigate possible flux density variability we retrieved Very
Large Array (VLA)
archival data for both J1548$+$3511 and PKS\,2004$-$447 (see Table
\ref{vla_flux}). 
Data were reduced following the standard procedure implemented 
in the \texttt{AIPS} package. Primary calibrators are 3C\,286 and
3C\,48 and the uncertainties on the amplitude calibration
are between 3 per cent and 5 per cent.
VLA images were obtained after a few phase-only
self-calibration iterations. Both sources are unresolved on arcsecond scale. \\
The errors on the VLA flux densities are dominated by the uncertainties
on the amplitude
calibration, being 
$\sigma_{m} \sim$0.1 mJy beam$^{-1}$. \\

\begin{table*}
\caption{VLBA flux density of J1548$+$3511 and PKS\,2004$-$447. Column 1:
  source name; Column 2: source component; Columns 3, 4, 5, and 6: flux
  density at 1.4, 5, 8.4, and 15 GHz, respectively.}
\begin{center}
\begin{tabular}{cccccc}
\hline
Freq.&Comp.& $S_{\rm 1.4}$&$S_{\rm 5}$&$S_{\rm 8.4}$&$S_{\rm 15}$\\
  GHz&     & mJy & mJy & mJy & mJy\\
\hline
J1548$+$3511   & C & - & 12.9$\pm$0.6 & 16.0$\pm$0.8 & 20.8$\pm$1.5 \\
               & J & - &  3.8$\pm$0.2 & 0.8$\pm$0.1  & - \\
               & J0& - & -  & 1.5$\pm$0.1  & - \\
               & J1& - & 9.7$\pm$0.6 & 5.1$\pm$0.3  & - \\
               &Ext& - & 11.0$\pm$0.8 & 4.4$\pm$0.5  & - \\
               &Tot& - & 37.4$\pm$2.0 & 27.8$\pm$1.4 &20.8$\pm$1.5\\
PKS\,2004$-$447& C &224$\pm$22& - & - & -\\
               & J &100$\pm$10& - & - & -\\
               & J1&210$\pm$21& - & - & -\\
               &Tot&534$\pm$53& - & - & -\\
\hline
\end{tabular}
\end{center}
\label{flux_vlba}
\end{table*}

\begin{table*}
\caption{Archival VLA data. Column 1: source name; Column 2: Observing
frequency; Columns 3, 4: date and project code of the observation;
Column 5: beam size; Column 6: observing time; Column 7: flux density.} 
\begin{center}
\begin{tabular}{ccccccc}
\hline
Source& Freq & Date & Code & Beam & Obs. time & Flux \\
      & GHz  &      &      & arcsec$^{2}$& min & mJy \\
\hline
J1548+3511 & 1.4 & 1999-04-13 & AL485 &49.31$\times$46.17  & 1 & 136$\pm$4 \\  
           & 1.4 & 2003-09-05 & AL595 &1.46$\times$1.17& 1.2 &124$\pm$6 \\  
           & 4.8 & 1994-05-07 & AK360 &0.74$\times$0.39& 1.0 & 74$\pm$2 \\  
           & 8.4 & 1995-08-14 & AM484 &0.24$\times$0.22& 0.6 & 76$\pm$2 \\  
           & 8.4 & 1999-04-13 & AL485 &8.46$\times$7.65& 0.8 & 54$\pm$2 \\  
PKS\,2004-447 & 8.4&1988-12-23& AP001 &1.29$\times$0.19& 1 & 330$\pm$17 \\ 
              & 8.4&1995-07-15& AK394 &1.15$\times$0.19& 4.5 & 270$\pm$11 \\  
              & 8.4&2005-03-16 &AK583 &4.63$\times$0.56& 3 & 219$\pm$11 \\ 
              & 8.4&2005-04-21 &AK583 &3.65$\times$0.60& 6 & 440$\pm$18 \\ 
\hline
\end{tabular}
\end{center}
\label{vla_flux}
\end{table*}

\section{{\it Fermi}-LAT Data: Selection and Analysis}
\label{FermiData}

The LAT on board the {\it Fermi} satellite is a $\gamma$-ray telescope operating from $20$\,MeV to $>300$\,GeV, with a large peak effective area ($\sim$ $8000$\,cm$^2$ for $1$\,GeV photons), an energy resolution typically $\sim$10 per cent, and a field of view of about 2.4\,sr with single-photon angular resolution (68 per cent containment radius) of 0\fdg6 at $E = 1$ GeV on-axis. Details about the LAT are given by \citet{atwood09}.

\begin{figure}
\begin{center}
\includegraphics{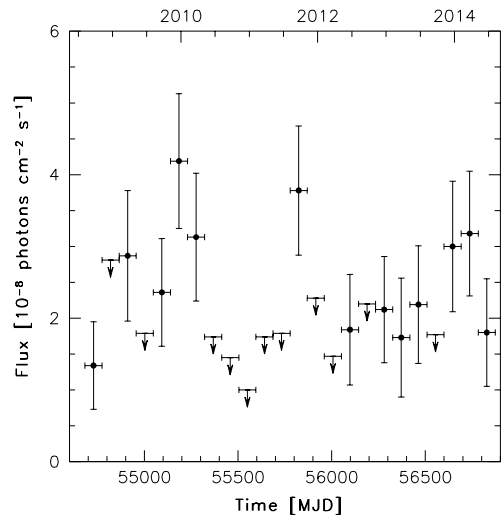}
\vspace{6.5cm}
\caption{LAT light curve of PKS\,2004$-$447 in the 0.1--100 GeV
  energy range obtained from 2008 August 4 to 2014 August 4 (MJD
  54682--56873) with 3-month time bins. Downward arrows represent
  2$\sigma$ upper limits.} 
\label{LAT_2004}
\end{center}
\end{figure}

The LAT data reported in this paper for PKS\,2004$-$447 and
J1548$+$3511 were collected over the first 6 years of {\it Fermi}
operation, from 2008 August 4 (MJD 54682) to 2014 August 4 (MJD
56873). 
During this time, the {\it Fermi} observatory operated almost entirely
in survey mode. The analysis was performed with the
\texttt{ScienceTools} software package version v9r33p0\footnote{\url{http://fermi.gsfc.nasa.gov/ssc/data/analysis/software/}}. Only events
belonging to the `Source' class were used. The time intervals when the
rocking angle of the LAT was greater than 52$^{\circ}$ were
rejected. In addition, a cut on the zenith angle ($< 100^{\circ}$) was
applied to reduce contamination from the Earth limb $\gamma$ rays,
which are produced by cosmic rays interacting with the upper
atmosphere. The spectral analysis was performed with the instrument
response functions \texttt{P7REP\_SOURCE\_V15} using an unbinned
maximum-likelihood method implemented in the tool 
\texttt{gtlike}. Isotropic (`iso\_source\_v05.txt') 
and Galactic diffuse emission (`gll\_iem\_v05\_rev1.fit') components
were used to model the 
background\footnote{\url{http://fermi.gsfc.nasa.gov/ssc/data/access/lat/Background\\Models.html}}. 
The normalizations of both components were allowed to vary freely
during the spectral fitting. \\ 
We analysed a region of interest of $10^{\circ}$ radius centred at the
location of our two targets. We evaluated the significance of the
$\gamma$-ray signal from the source by means of the maximum-likelihood
test statistic ${\rm TS} = 2 \times({\rm log}L_{1} - {\rm log}L_{0}$), where $L$ is the
likelihood of the data given the model with ($L_1$) or without ($L_0$)
a point source at the position of our target
\citep[e.g.,][]{mattox96}. The source model used in \texttt{gtlike}
includes all the point sources from the third {\it Fermi}-LAT
catalogue \citep[3FGL;][]{acero15} that fall within $15^{\circ}$ of
the target. The spectra of these sources were parametrized by
power-law, log-parabola, or exponential cut-off power-law model, as
in the 3FGL catalogue.\\ 
A first maximum-likelihood analysis was performed to remove from the
model the sources having ${\rm TS} < 10$ and/or a predicted number of counts
based on the fitted model $N_{\rm pred} < 1 $. A second
maximum-likelihood analysis was performed on the updated source
model. In the fitting procedure, the 
normalization factors and the photon indices of the sources lying
within 10$^{\circ}$ of the target were left as free parameters. For
the sources located between 10$^{\circ}$ and 15$^{\circ}$ from the
target, we kept the normalization and the photon index fixed to the
values from the 3FGL catalogue.\\
For PKS\,2004$-$447, the fit with a power-law model to the data integrated over
72 months of {\it Fermi}-LAT 
operation in the 0.1--100 GeV energy range results in
${\rm TS} = 164$, with an average flux of ($1.59 \pm 0.16$)
$\times$10$^{-8}$ ph cm$^{-2}$ s$^{-1}$, and a photon index $\Gamma_{\gamma} =
2.39 \pm 0.06$, corresponding to an energy flux of
  (1.09$\pm$0.11)$\times$10$^{-12}$ erg cm$^{-2}$ s$^{-1}$. Figure \ref{LAT_2004} shows the $\gamma$-ray light curve of 
PKS\,2004$-$447 for the period 2008 August 4--2014 August 4 using 3-month 
time bins. For each time bin, the photon index was frozen to the value
resulting from the likelihood analysis over the whole period. 
The systematic uncertainty in the effective area \citep{ackermann12} amounts to
10 per cent below 100 MeV, decreasing linearly with the logarithm of energy to
5 per cent between 316 MeV and 10 GeV, and increasing linearly with
the logarithm 
of energy up to 15 per cent at 1 TeV
\footnote{http://fermi.gsfc.nasa.gov/ssc/data/analysis/LAT\_caveats.html}. Statistical
errors dominate over the systematics uncertainty. All errors relative to $\gamma$-ray data reported throughout the paper are statistical only. \\ 
For J1548$+$3511, the fit with a power-law model over 6 years of {\it Fermi}-LAT
operation results in ${\rm TS} = 1$. The 2$\sigma$ upper limit is
3.35$\times$10$^{-8}$ ph cm$^{-2}$ s$^{-1}$ in the 0.1--100 GeV
energy range (assuming a photon 
index $\Gamma_{\gamma} = 2.4)$, corresponding to an energy flux
  $<$5.3$\times$10$^{-11}$ erg cm$^{-2}$ s$^{-1}$.\\

\section{Swift data: observations and analysis}
\label{SwiftData}

The {\em Swift} satellite \citep{gehrels04} performed twenty seven
observations of 
PKS\,2004$-$447 between 2011 May 15 and 2014 March 16. 
J1548+3511 has not been observed by {\it Swift} so far.\\ 
The observations of PKS\,2004$-$447
were carried 
out with all three instruments on board: the X-ray Telescope
\citep[XRT;][0.2--10.0 
keV]{burrows05}, the Ultraviolet/Optical Telescope \citep[UVOT;][170--600
nm]{roming05} and the Burst Alert Telescope \citep[BAT;][15--150 keV]{barthelmy05}.
%
The source was not present in the {\em Swift} BAT
70-month hard X-ray catalogue \citep{baumgartner13}.

The XRT data of PKS\,2004$-$447 were processed with standard procedures (\texttt{xrtpipeline v0.13.0}),
filtering, and screening criteria using the \texttt{HEAsoft}
package (v6.15). The data were collected in photon counting mode for all
the observations. The source count rate was low ($< 0.5$ counts s$^{-1}$);
thus pile-up correction was not required. Source events were extracted from a
circular region with a radius of 20 pixels (1 ${\rm pixel} = 2.36$ arcsec), while
background events were extracted from a circular region with radius of 50
pixels away from the source region and from other bright sources. Ancillary
response files were generated with \texttt{xrtmkarf}, and account for
different extraction regions, vignetting and point-spread function
corrections. We used the spectral redistribution matrices in the Calibration
data base maintained by HEASARC\footnote{\url{http://heasarc.nasa.gov/}}. Short
observations
performed during the same month were summed in order to have enough statistics
to obtain a good spectral fit.
The spectra with low numbers of photons collected ($< 200$ counts) were rebinned
with a minimum of 1 count per
bin and the Cash statistic \citep{cash79} was used. We fitted the spectra with an
absorbed power-law using the photoelectric absorption model
\texttt{tbabs} \citep{wilms00}, with a neutral hydrogen column density fixed to its
Galactic value \citep[3.17$\times$10$^{20}$cm$^{-2}$;][]{kalberla05}. The fit
results are reported in Table~\ref{XRT}.\\ 

UVOT data of PKS\,2004$-$447 in the $v$, $b$, $u$, $w1$, $m2$, and
$w2$ filters were reduced with the task \texttt{uvotsource} included
in the 
\texttt{HEAsoft} package v6.15 and the 20130118 CALDB-UVOTA
release. We extracted the source counts from a circle with 5 arcsec radius
centred on the source and the background counts from a circle with 10 arcsec
radius in a nearby empty region. The observed magnitudes are reported in
Table~\ref{uvot}. We converted the magnitudes into de-reddened flux
densities by using the E(B-V) value of 0.029 from \citet{schlafly11}, the
extinction laws by \citet{cardelli89} and the magnitude-flux calibrations by
\citet{bessell98} and \citet{breeveld11}. \\

\begin{table*}
\caption{Log and fitting results of {\it Swift}-XRT observations of PKS\,2004$-$447
using a power-law model with an HI column density fixed to the
  Galactic value in the direction of the source. $^{a}$Unabsorbed flux.}
\begin{center}
\begin{tabular}{ccccc}
\hline 
\multicolumn{1}{c}{Date} &
\multicolumn{1}{c}{Date} &
\multicolumn{1}{c}{Net exposure time} &
\multicolumn{1}{c}{Photon index} &
\multicolumn{1}{c}{Flux 0.3--10 keV$^{a}$} \\
\multicolumn{1}{c}{MJD} &
\multicolumn{1}{c}{UT} &
\multicolumn{1}{c}{s} &
\multicolumn{1}{c}{$\Gamma_{\rm X}$} &
\multicolumn{1}{c}{10$^{-13}$ erg cm$^{-2}$ s$^{-1}$} \\
\hline
55686 & 2011-05-05 & 6778 & 1.65 $\pm$ 0.25 & $5.6 \pm 0.9$ \\
55756/55761/55765 & 2011-07-14/19/23 & 7101 & 1.62 $\pm$ 0.29 & $5.5 \pm 0.7$ \\
55783/55786 & 2011-08-10/13 & 2258 & 1.15 $\pm$ 0.49 & $6.4 \pm 2.1$ \\
55809/55821 & 2011-09-05/17 & 8239 & 1.38 $\pm$ 0.19 & $11.8 \pm 1.3$ \\
55880 & 2011-11-15 & 7095 & 2.02 $\pm$ 0.30 & $4.4 \pm 0.7$ \\
56000 & 2012-03-14 & 7306 & 1.52 $\pm$ 0.30 & $4.5 \pm 0.8$ \\
56111/56120 & 2012-07-3/12 & 7120 & 1.65 $\pm$ 0.24 & $6.5 \pm 0.9$ \\
56182/56192/56200 & 2012-09-12/22/30 & 17016 & 1.63 $\pm$ 0.18 & $5.9 \pm 0.6$ \\
56480/56487 & 2013-07-07/14 & 22963 & 1.42 $\pm$ 0.12 & $10.4 \pm 0.7$ \\
56562 & 2013-09-27 & 8361 & 1.45 $\pm$ 0.20 & $14.6 \pm 1.5$ \\
56578/56585/56592/56594 & 2013-10-13/20/27/29 & 12699 & 1.41 $\pm$ 0.24 & $8.7 \pm
1.2$ \\
56599/56618 & 2013-11-03/19 & 8241 & 1.58 $\pm$ 0.18 & $6.8 \pm 0.6$ \\
56619 & 2013-11-20 & 12179 & 1.43 $\pm$ 0.16 & $15.5 \pm 1.2$ \\
56730 & 2014-03-14 & 7507 & 1.68 $\pm$ 0.26 & $6.7 \pm 1.0$ \\
56732 & 2014-03-16 & 9642 & 1.48 $\pm$ 0.19 & $9.0 \pm 1.0$ \\
\hline
\end{tabular}
\end{center}
\label{XRT}
\end{table*}

\begin{table*}
\caption{Results of the {\it Swift}-UVOT observations of PKS\,2004$-$447 in magnitudes.}
\begin{center}
\begin{tabular}{cccccccc}
\hline 
\multicolumn{1}{c}{Date (MJD)} &
\multicolumn{1}{c}{Date (UT)} &
\multicolumn{1}{c}{$v$} &
\multicolumn{1}{c}{$b$} &
\multicolumn{1}{c}{$u$} &
\multicolumn{1}{c}{$w1$} &
\multicolumn{1}{c}{$m2$} &
\multicolumn{1}{c}{$w2$} \\
\hline
55686 & 2011-05-05 & 18.48$\pm$0.18 & 20.08$\pm$0.28 & 19.17$\pm$0.19 &
19.77$\pm$0.32 & $>$ 19.86 & $>$ 20.19 \\
55756 & 2011-07-14 & $>$ 18.98 & 19.69$\pm$0.30 & 19.31$\pm$0.28 & 19.29$\pm$0.29 &
$>$ 19.40 & $>$ 19.85 \\
55761 & 2011-07-19 & 18.85$\pm$0.35 & -- & $>$ 18.89 & $>$ 19.35 & $>$ 19.19 & $>$
19.61 \\ 
55765 & 2011-07-23 & $>$ 19.33 & 20.25$\pm$0.36 & 19.25$\pm$0.20 & 19.53$\pm$0.28 &
$>$ 19.70 & $>$ 20.01 \\
55783 & 2011-08-10 & -- & -- & -- & 19.41$\pm$0.18 & -- & -- \\
55786 & 2011-08-13 & 18.48$\pm$0.28 & 19.56$\pm$0.31 & 19.29$\pm$0.33 &
19.67$\pm$0.35 & $>$ 19.87 & 20.37$\pm$0.35 \\
55809 & 2011-09-05 & $>$18.24 & $>$19.12 & 18.59$\pm$0.32 & $>$ 19.06 & $>$ 18.98 &
19.74$\pm$0.34 \\
55821 & 2011-09-17 & 18.28$\pm$0.25 & 18.99$\pm$0.16 & 18.77$\pm$0.18 &
18.84$\pm$0.24 & $>$ 19.30 & $>$ 19.50 \\
55880 & 2011-11-15 & $>$ 18.61 & 19.63$\pm$0.32 & $>$ 19.25 & $>$ 19.33 & $>$ 19.40
& $>$ 19.52 \\
56000 & 2012-03-14 & $>$ 18.94 & $>$ 19.76 & $>$ 19.41 & $>$ 19.59 & $>$ 19.70 &
19.55$\pm$0.25 \\
56111 & 2012-07-03 & -- & -- & -- & 19.48$\pm$0.11 & -- & -- \\
56120 & 2012-07-12 & -- & -- & -- & -- & 19.94$\pm$0.35 & -- \\
56182 & 2012-09-12 & 18.45$\pm$0.24 & 19.67$\pm$0.30 & 18.91$\pm$0.22 &
19.46$\pm$0.28  & 19.62$\pm$0.14 & 20.00$\pm$0.23 \\
56192 & 2012-09-22 & $>$ 18.90 & 19.36$\pm$0.23 & 18.72$\pm$0.19 & 19.37$\pm$0.32 &
19.70$\pm$0.27 & 20.27$\pm$0.30 \\
56200 & 2012-09-30 & -- & -- & 19.10$\pm$0.09 & -- & -- & -- \\
56480 & 2013-07-07 & -- & -- & 18.87$\pm$0.09 & -- & -- & 19.91$\pm$0.20 \\
56487 & 2013-07-14 & -- & -- & 18.56$\pm$0.07 & 19.36$\pm$0.10 & -- & -- \\
56562 & 2013-09-27 & -- & -- & -- & 19.03$\pm$0.09 & 19.28$\pm$0.11 & -- \\
56578 & 2013-10-13 & -- & -- & -- & -- & 19.81$\pm$0.23 & -- \\
56585 & 2013-10-20 & -- & -- & -- & -- & -- & 19.77$\pm$0.15 \\
56592 & 2013-10-27 & -- & -- & 18.69$\pm$0.21 & -- & -- & 19.54$\pm$0.25 \\
56594 & 2013-10-29 & -- & -- & -- & -- & 19.91$\pm$0.23 & -- \\ 
56599 & 2013-11-03 & -- & -- & -- & 19.28$\pm$0.27 & -- & -- \\
56618 & 2013-11-19 & -- & -- & -- & 18.85$\pm$0.15 & -- & -- \\
56619 & 2013-11-20 & -- & -- & 18.30$\pm$0.06 & -- & -- & -- \\
56730 & 2014-03-14 & -- & -- & -- & 19.67$\pm$0.16 & $>$ 19.56 & -- \\
56732 & 2014-03-16 & -- & -- & 18.59$\pm$0.11 & -- & -- & -- \\
\hline
\end{tabular}
\end{center}
\label{uvot}
\end{table*}

\section{{\it XMM-Newton} data: observations and analysis}
\label{XMM}

\subsection{EPIC Observations and data reduction}
\label{observations}

J1548$+$3511 was observed by {\em XMM-Newton} \citep{jansen01} on 2011
August 8 and 20 for a total observing time of 28 ks in both cases
(observation IDs 0674320301 and
0674320401). For this analysis we focus on the data from the EPIC pn, which was
operated in the Full Window mode, with net exposure
times of 19.4~ks and 22.9~ks for the two observations. \\
{\em XMM-Newton} observed PKS\,2004$-$447 on 2012 May 1 and
October 18 for a total duration of 36 ks in both cases (observation IDs
0694530101 and 0694530201). The EPIC pn and the EPIC MOS cameras (MOS1 and MOS2) were operated in the full
frame mode. The first observation has net exposure times of 18, 20, and 20~ks for the pn, MOS1
and MOS2, respectively; the second observation has net exposure times of 29,
35, and 35~ks. Differently from the first observation, in the second
observation there is a
significant increase of counts ($\sim$70 per cent) adding the data
collected by MOS1 and MOS2 to the pn data. Therefore,
for the first observation of PKS\,2004$-$447 we focus on the data
from the EPIC pn only, while for the second observation both the pn and MOS data
were analysed.\\
The data were reduced using the {\em XMM-Newton} Science Analysis
System ({\small SAS v13.5.0}), applying standard event selection and
filtering\footnote{\url{http://xmm.esac.esa.int/external/xmm_user_support/documentation/sas_usg/USG/}}. Neither observation was affected by background
flaring. The source spectra were extracted from a circular region of
radius 30 arcsec centred on the source, and the background from a
nearby region on the same chip. To allow for $\chi^2$ fitting the
spectra were binned to contain at least 20 counts per bin. All errors
are quoted at the 90 per cent confidence level for the parameter of
interest (corresponding to $\Delta \chi^2 = 2.7$).

\subsection{X-ray spectral analysis}

\subsubsection{J1548+3511}

\begin{figure*}
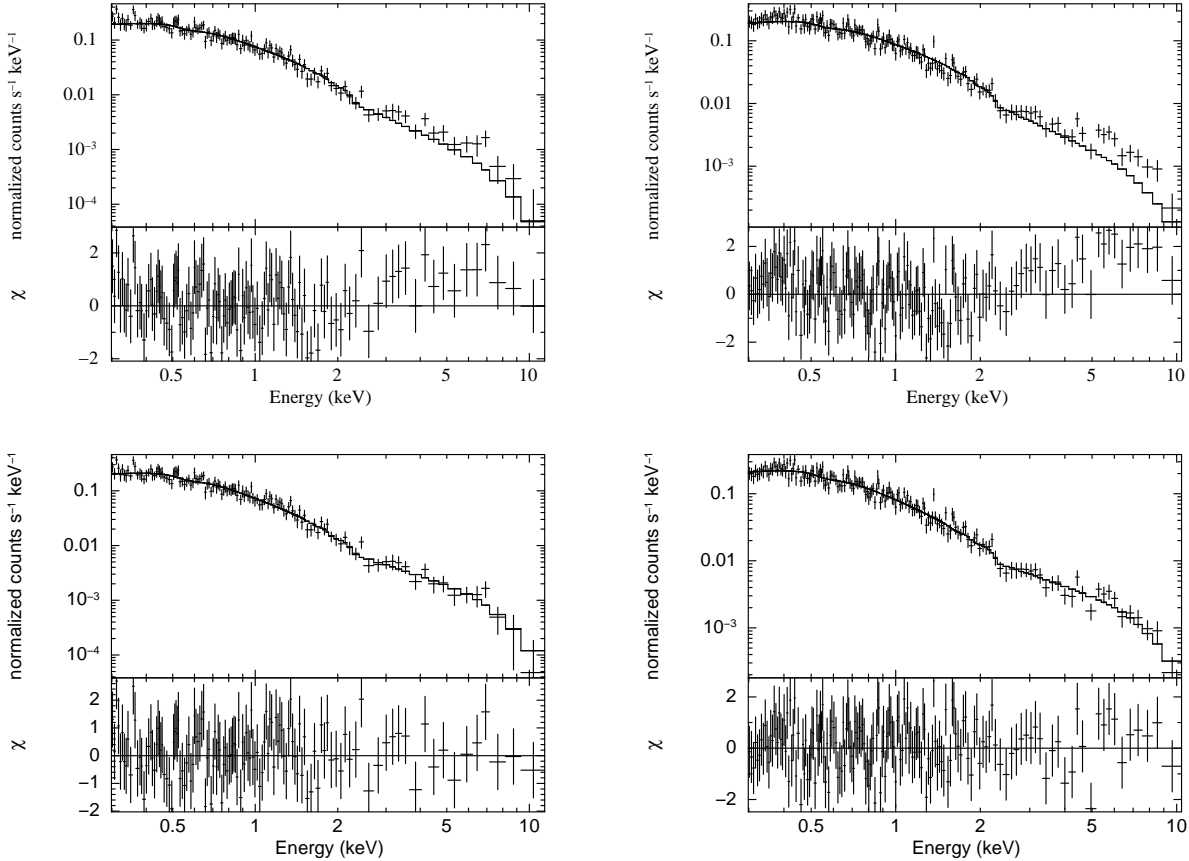

\begin{center}
\includegraphics{pow_obs1.eps}
\includegraphics{pow_obs2.eps}
\includegraphics{bknpow_obs1.eps}
\includegraphics{bknpow_obs2.eps}
\vspace{12cm}
\caption{EPIC pn spectra and residuals of J1548+3511 fitted with a power law (upper panel) and a broken power-law (lower panel). The first observation is shown to the left and the second one to the right.  }
\label{j1548_fits}
\end{center}
\end{figure*}

The spectral fits were performed over the 0.3--10~keV energy range
using  XSPEC v.12.8.1. Although we present only the fits to the EPIC
pn, the results were cross-checked for consistency with the MOS
spectra. Galactic absorption of \(2.19\times 10^{20}~\rm{cm^{-2}}\)
was included in all fits using the {\it tbabs} model. In the two weeks
between the observations the unabsorbed flux increased from
$F_{\rm 0.3-10~keV} = (4.2 \pm 0.2) \times 10^{-13} {\rm erg\ cm^{-2}\
  s^{-1}}$ to $F_{\rm 0.3-10~keV} = (5.3 \pm 0.2) \times
10^{-13} {\rm erg\ cm^{-2}\ s^{-1}}$. No strong variability was seen
during the individual observations.\\  
The results of fitting a single power law and a broken power-law to
the two spectra are summarised in Table~\ref{j1548_table} and shown in
Fig.~\ref{j1548_fits}. While the single power-law model leaves
positive residuals above 3 keV, the broken power-law is a good fit to
the data. The improvement between the models is more significant in
the second observation,  when the source was brighter. In this model
the spectral shape changes from a soft slope of $\Gamma_1 =
2.6\pm0.1$ ($2.5\pm0.1$) below $E_{\rm{break}} =
2.0^{+0.8}_{-0.5}~\rm{keV}$  ($1.7\pm 0.2 ~\rm{keV}$) to $\Gamma_2 =
1.9^{+0.3}_{-0.4}$ ($\Gamma_{2} = 1.7\pm 0.2$) above $E_{\rm{break}}$
for the first (second) observation. 
%
%
If we add to the models another neutral absorber at
the redshift of the source the fits do not improve, showing that no
intrinsic absorption is required. Furthermore, there is no detection
of an Fe line, with 90 per cent upper limits on a narrow line at 6.4 keV of
$EW< 0.26$ keV and $EW< 0.13$ keV for the first and second
observation, respectively. Unfortunately, the data quality is not
sufficient to obtain meaningful constraints on more complex models for
the emission.

\begin{table*}
    \caption{Fits to the 0.3--10~keV EPIC~pn spectra of
      J1548+3511. Galactic absorption was included in all fits.}
\begin{center}
    \begin{tabular}{llll}
    \hline \\ [-4pt]
    Model  & Parameter  & Value Obs 1 & Value Obs 2 \\
    \\
\hline\\[-6pt]
    Power law & \(\Gamma\) & \(2.47\pm0.06\) &  \(2.31\pm0.05\) \\
          & Norm  & \(7.8\pm 0.2 \times10^{-5}\) & \(9.0\pm 0.2 \times 10^{-5}\) \\ 
          & \(\chi^2\)/d.o.f.  & 129/129 & 210/164 \\ 
          \hline 
    Broken power-law & \(\Gamma_1\) & \(2.57^{+0.09}_{-0.08}\) &  \(2.49^{+0.09}_{-0.08}\)\\ 
          & \(E_{\rm{break}}~(\rm{keV})\) & \(2.0^{+0.8}_{-0.5}\)  &   \(1.7^{+0.5}_{-0.3}\)\\ 
          & \(\Gamma_2\) & \(1.9^{+0.3}_{-0.4}\)  & \(1.7\pm 0.2\) \\ 
          & Norm  & \(7.5^{+0.3}_{-0.4}\times10^{-5}\)  & \(8.4^{+0.3}_{-0.4}\times10^{-5}\)  \\ 
          & \(\chi^2\)/d.o.f.  & 110/127 & 154/162 \\ 
          \hline    
    \end{tabular}
\end{center}
\label{j1548_table}     
\end{table*}


\subsubsection{PKS\,2004$-$447}

\begin{figure}
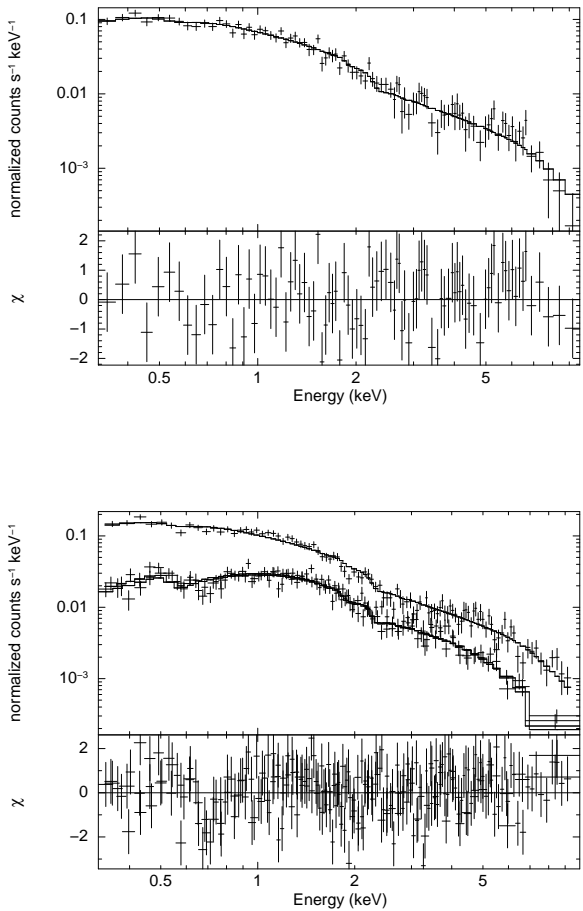

\begin{center}
\includegraphics{2004_I.ps}
\includegraphics{2004_II.ps}
\vspace{13cm}
\caption{EPIC spectra and residuals of PKS\,2004$-$447 fitted with a power law. For the first observation (upper panel) only EPIC pn data were considered, while for the second (bottom panel) EPIC pn, MOS1, and MOS2 were considered in the fit.}
\label{2004_fits}
\end{center}
\end{figure}

As for J1548$+$3511, the spectral fits were performed over the
0.3$-$10~keV energy range using XSPEC 
v.12.8.1. Although for the first observation we present only the fits
to the EPIC pn, the results were cross-checked for consistency with the MOS
spectra. Galactic absorption of \(3.17\times 10^{20}~\rm{cm^{-2}}\) was
included in all fits using the {\it tbabs} model. In the five months between
the observations the unabsorbed
flux increased from  $F_{0.3-10~\rm{keV}} = (4.5~\pm 0.1) \times
10^{-13}~\rm{erg\ cm^{-2}\ s^{-1}}$ to $F_{0.3-10~\rm{keV}} = (6.8~\pm 0.2) \times 10^{-13}~\rm{erg\ cm^{-2}\ s^{-1}}$. No strong variability was seen
during the individual observations. \\
The results of fitting a single power law and a broken power-law to the two
spectra are summarised in Table~\ref{2004_XMM}. 
%
%
No significant soft X-ray excess is observed below
2 keV. A simple power-law model (Fig.~\ref{2004_fits}) is sufficient
to describe the data of the first observation. For the second
observation the power-law is not an optimal fit ($\chi^{2} = 273/241$,
Table \ref{2004_XMM}), but
no significant
improvement was obtained using a broken power-law model. 
We note a dip in the residuals of the second observation at $\sim$0.7
keV. Adding to the model an absorption edge, the fit is slightly
better ($\chi^{2} = 263/239$) with a threshold energy $E_{\rm
  c} = 0.77^{+0.05}_{-0.11}$ keV, compatible with the O$_{\rm VII}$
absorption, and a maximum optical depth
$\tau = 0.21^{+0.14}_{-0.11}$. However, observations with better
statistics are required to confirm this feature.\\ 
In both observations, the photon index is
$\Gamma_{\rm X}$ $\sim$1.7, consistent with a jet emission component.
Similarly to J1548+3511, if we add to the models another neutral absorber at the
redshift of the source the fits do not improve,
showing that no intrinsic absorption is required. Moreover, there is
no detection of an Fe line, with 90 per cent upper limits on a narrow
line at 6.4~keV of  $EW< 0.12$~keV and $EW< 0.05$~keV for the first
and second observation, respectively.\\

\begin{table*}
    \caption{Fits to the 0.3--10~keV EPIC spectra of PKS\,2004$-$447. Galactic absorption was included in all fits.}
\begin{center}
    \begin{tabular}{llll}
    \hline \\ [-4pt]
    Model  & Parameter  & Value Obs 1 & Value Obs 2 \\
    \\
\hline\\[-6pt]
    Power law & \(\Gamma\) & \(1.72\pm0.05\) &  \(1.69\pm0.03\) \\
          & Norm  & \(7.0\pm 0.2 \times10^{-5}\) & \(10.0\pm 0.3 \times 10^{-5}\) \\ 
          & \(\chi^2\)/d.o.f.  & 97/102 & 273/241 \\ 
          \hline 
    Broken power-law & \(\Gamma_1\) & \(1.76\pm 0.07\) &  \(1.71\pm 0.03\)\\ 
          & \(E_{\rm{break}}~(\rm{keV})\) & \(2.0^{+2.0}_{-0.9}\)  &   \(3.3^{+1.2}_{-0.6}\)\\ 
          & \(\Gamma_2\) & \(1.63\pm 0.19\)  & \(1.51\pm 0.14\) \\ 
          & Norm  & \(7.0\times10^{-5}\)  & \(10.4\pm 0.3\times10^{-5}\)  \\ 
          & \(\chi^2\)/d.o.f.  & 96/100 & 268/239 \\ 
          \hline    
    \end{tabular}
\end{center}
          \label{2004_XMM}
\end{table*}

\subsection{Optical Monitor data}

The Optical Monitor \citep[OM;][]{mason01} on board {\it XMM-Newton} is a 30
cm telescope carrying six optical/UV filters, and two grisms.  We used the SAS
task \texttt{omichain} to reduce the data and the tasks \texttt{omsource} and \texttt{omphotom} to
derive the source magnitude. \\
The OM was operated during all the observations described in Section 5.1.
J1548$+$3511 was observed twice in optical and UV bands.
Average observed magnitudes for J1548$+$3511 are
reported in Table \ref{OM_1548}. No significant change of activity was
observed for J1548$+$3511 between the two OM observations. PKS\,2004$-$447 was
observed by OM in $u$ band, with observed magnitudes $u = 19.20\pm0.03$ and
$u = 18.92\pm0.03$ for 2012 May 1 and October 18,
respectively. Therefore, on a half-year time-scale a difference
of $\sim$0.3 mag was observed in optical for PKS\,2004$-$447.

\begin{table*}
\caption{Results of the {\em XMM}-OM observations of J1548$+$3511 in magnitudes.}
\begin{center}
\begin{tabular}{cccccccc}
\hline 
\multicolumn{1}{c}{Date (MJD)} &
\multicolumn{1}{c}{Date (UT)} &
\multicolumn{1}{c}{$v$} &
\multicolumn{1}{c}{$b$} &
\multicolumn{1}{c}{$u$} &
\multicolumn{1}{c}{$w1$} &
\multicolumn{1}{c}{$m2$} &
\multicolumn{1}{c}{$w2$} \\
\hline
55781 & 2011-08-08 &  18.10$\pm$0.10 & 18.30$\pm$0.03 & 17.37$\pm$0.02 & 16.99$\pm$0.02 & 16.62$\pm$0.08 & 16.77$\pm$0.07 \\
55793 & 2011-08-20 &  17.99$\pm$0.07 & -- & 17.28$\pm$0.02 & 16.99$\pm$0.02 & 16.68$\pm$0.08 & 16.78$\pm$0.09 \\
\hline
\end{tabular}
\end{center}
\label{OM_1548}
\end{table*}

\section{Results}

\subsection{The NLSy1 J1548$+$3511}

\subsubsection{Radio properties}

In the radio band, the NLSy1 J1548$+$3511 shows 
a core-jet structure with a total angular
size of about 70 mas corresponding to a linear size of 420
pc (Fig. \ref{1548_vlba}). Component C accounts for the majority of
the VLBA flux density, from 
about 34 per cent at 5 GHz up to 100 per cent at 15 GHz. Its inverted
spectrum ($\alpha \sim -0.4$; $S_{\nu} \propto \nu^{- \alpha}$)
indicates that is the 
source core. From the core region a one-sided jet emerges with a
position angle of 
about 10$^{\circ}$ and bends to the east (position angle of about 
30$^{\circ}$) at a distance of $\sim$40 mas (240 pc) from the core,
where the component J1 is observed. The
higher resolution provided by 8.4-GHz data allows us to resolve the
innermost part of the jet into
two compact components (labelled J and J0 in Fig. \ref{1548_vlba}), 
which are likely jet knots. An extended low-surface brightness
structure (labelled Ext in Figs. \ref{1548_vlba} and
\ref{1548_natural}) is observed at $\sim$60 mas (360 pc) from the core.
%
%
At 15 GHz only the core component is detected.\\
The source is unresolved on the arcsecond scale sampled by VLA
images. The similar flux density ($\sim$141 mJy) reported in the NVSS
\citep{condon98} and the FIRST \citep{becker95} indicates that no
extended emission on arcsecond scale is present. On the other hand,
VLBA observations at 5 and 8.4 GHz could recover only about 50 per
cent or less of the VLA flux density. This may be due to a combination
of both variability and the presence of extended jet structure that cannot
be imaged by the short baselines of the VLBA. The spectral index
computed using VLA data between 1.4 and 8.4 GHz results in a
moderately flat spectrum with $\alpha \sim 0.3$--0.4. However, the
spectral index values are strongly subject to the flux density
variability observed in this source.\\  
A small part of the flux density from the
extended low-surface brightness structure is recovered in the low
resolution image at 8.4 (Fig. \ref{1548_natural}), where the total flux density is $\sim$28
mJy instead of 21 mJy measured on the high-resolution image. No significant
difference between the flux density measured in low- and
high-resolution images is found at 15 GHz.\\ 
%
%
We investigated the source variability by the analysis of archival VLA
data. Although the data sets considered are not homogeneous
(i.e. different VLA configurations), the lack of extended
emission on arcsecond scale implies that variation in the flux
density should be intrinsic to the source and not related to the
lack of short baselines in the extended VLA configurations. Evidence
for intrinsic variability comes from observations at 8.4 GHz, where
the highest flux density was measured when the VLA was in the most
extended configuration (Table \ref{vla_flux}). \\
For each frequency we computed the variability index $V$ following
\citet{hovatta08}:

\begin{equation}
V = \frac{(S_{\rm max} - \sigma_{\rm max}) - (S_{\rm min} + \sigma_{\rm
min})}{(S_{\rm max} - \sigma_{\rm max}) + (S_{\rm min} + \sigma_{\rm
min})}
\label{var_index}
\end{equation}

\noindent where $S_{\rm max}$ and $S_{\rm min}$ are the maximum and
minimum flux density, whereas $\sigma_{\rm max}$ and $\sigma_{\rm
  min}$ are their associated errors, respectively. \\
In addition to the VLA flux density reported in Table \ref{vla_flux},
at 1.4 GHz we considered the values from the FIRST and NVSS, while at 5
GHz we considered the values reported in the 87GB catalogue \citep{gregory91}
and in the second MIT-Green Bank survey \citep{langston90}. VLBA
flux densities were not taken into account due to the possible missing flux from
extended jet structures on parsec scales. 
From Eq. \ref{var_index} we found that $V$ is
1 per cent, 11 per cent and 14 per cent, at 1.4, 5, and 8.4 GHz
respectively, indicating larger variability at higher frequencies as
usually found in blazars. The low variability ($V=0.01$) estimated at 1.4
GHz is comparable to the uncertainties.
We note that the flux density variability may be
underestimated due 
to the poor time sampling of the observations.\\

\subsubsection{X-ray properties}

In the two weeks between the observations the source brightened by
about 25 per cent.
The {\it XMM-Newton} spectra of J1548$+$3511 are well fitted by a
broken power-law model, with a possible Seyfert component 
below $\sim$2 keV and a jet component dominating at higher energies. The
low-energy component, with a steep photon index of $\sim$2.5, may be
associated with the soft X-ray excess.
%
%
On the other hand, the relatively hard X-ray spectrum above the energy
break ($\Gamma_{2} = 1.7$--1.9) may
suggest a significant contribution of inverse Compton radiation from a relativistic jet. \\
%
%

\subsubsection{Optical and UV properties}

No significant change of activity was observed for J1548$+$3511 in the optical
and UV bands between the two {\em XMM}-OM observations performed two weeks
apart (Table \ref{OM_1548}).

\subsection{The NLSy1 PKS\,2004$-$447}

\subsubsection{Radio properties}

The radio source PKS\,2004$-$447 has a core-jet structure with a total
angular size of about 40 mas, which corresponds to $\sim$150 pc at the
redshift of the source. The radio emission is
dominated by the source core, labelled C in Fig. \ref{2004_vlba},
which accounts for $\sim$42 per cent of the 
total flux density at 1.4 GHz. 
The jet structure emerges from the core component
with a position angle of about $-$90$^{\circ}$, then at 20 mas
($\sim$75 pc) it bends to a position angle of about
$-$60$^{\circ}$. The jet structure is resolved into two subcomponents, J
and J1, which are enshrouded by diffuse emission. The lack of
multifrequency observations does not allow us to study the spectral
index distribution across the source.\\
The source is unresolved in VLA images at 8.4 GHz, in agreement with
previous studies at arcsecond-scale resolution by \citet{gallo06}.\\
Archival VLA observations at 8.4 GHz point out some level of flux
density variability (Table \ref{vla_flux}). From Eq. \ref{var_index}
we computed the 
variability index for PKS\,2004$-$447, which turns out to be 27
per cent, consistent with the flux density variability
derived from the Ceduna observations at 6.65 GHz \citep{gallo06}. \\
%
%

\subsubsection{$\gamma$-ray properties}

During the first six years of {\it Fermi}-LAT observations, the
0.1--100 GeV averaged flux is
$\sim$1.6$\times$10$^{-8}$ ph cm$^{-2}$ s$^{-1}$. The LAT
light curve indicates variable $\gamma$-ray emission with flux ranging between
(1.3--4.2)$\times$10$^{-8}$ ph cm$^{-2}$ s$^{-1}$, interleaved by periods of low activity, when the
source is not detected by {\it Fermi}-LAT (Fig. \ref{LAT_2004}). 
No $\gamma$-ray
flares from this source have been
detected so far. \\

\subsubsection{X-ray properties}

\begin{figure*}
\begin{center}
\includegraphics{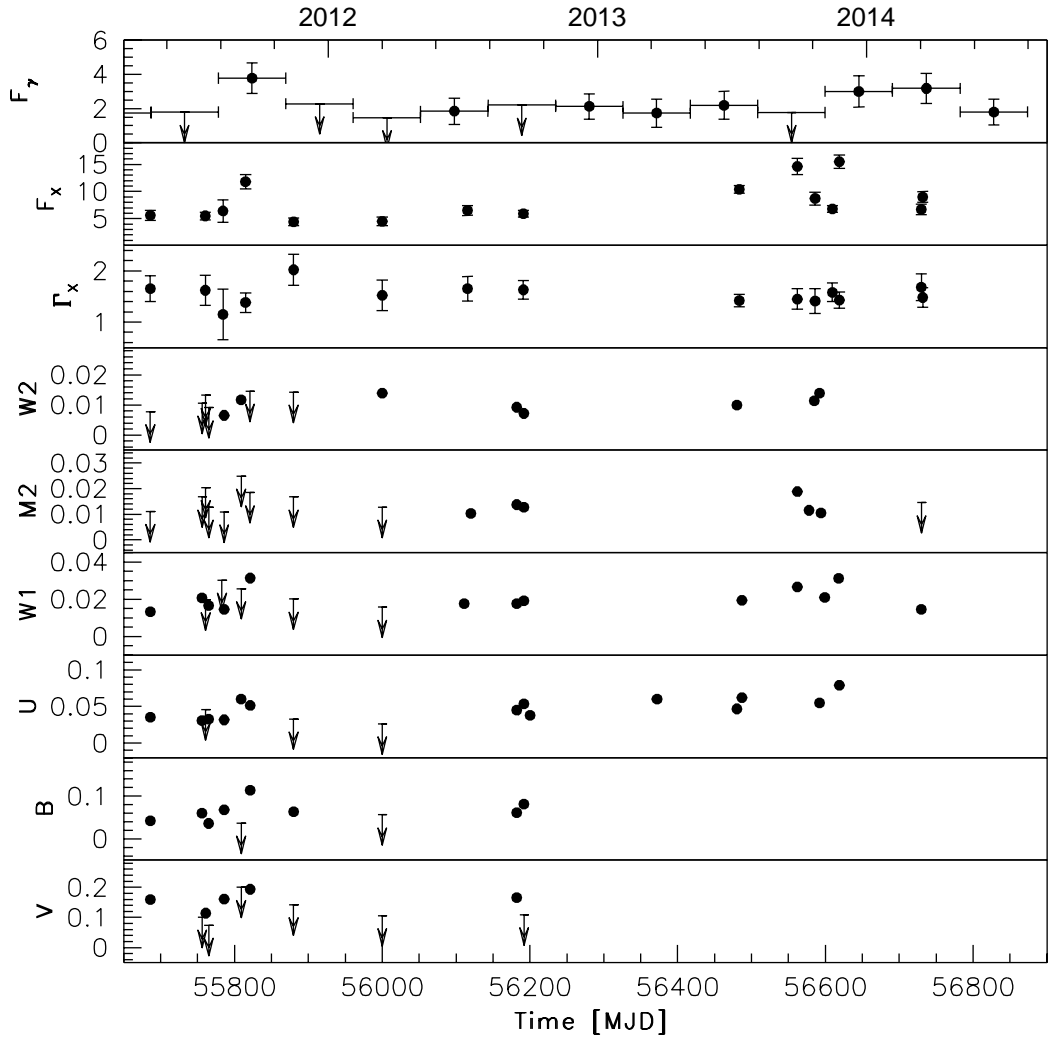}
\vspace{14.5cm}
\caption{Multifrequency light curve of PKS\,2004$-$447 for the period  between 2011 May and
2014 August. The LAT $\gamma$-ray light curve (in units of
  10$^{-8}$ ph cm$^{-2}$ s$^{-1}$ for $0.1< {\rm E}<100$ GeV; {\it upper panel}), the
{\em Swift} X-ray light curve (in units of 10$^{-13}$ erg
  cm$^{-2}$ s$^{-1}$ for $0.3< {\rm E}<10$ keV; {\it second panel from top}), the X-ray
photon index trend ({\em third panel}), the {\em Swift}-UVOT light curve in the
optical and UV filters ($v$, $b$, $u$, $w1$, $m2$, $w2$, in units of mJy;
{\it fourth to ninth panel}). Arrows refer to 2$\sigma$ and 3$\sigma$
upper limits 
on the source flux for LAT and {\em Swift}-UVOT measurements, respectively.}
\label{MWL}
\end{center}
\end{figure*}

The X-ray light curve collected by {\it Swift}-XRT indicates significant flux variability between
2011 and 2014, ranging between (5--16)$\times$10$^{-13}$ erg cm$^{-2}$
s$^{-1}$ (Fig. \ref{MWL}). 
In 2011 September the increase of the X-ray flux occurs
when the $\gamma$-ray emission is in a maximum, suggesting a
possible correlation between the emission in these two energy
bands. On the other hand, between 2013 September and November 
the X-ray light curve shows two high-activity episodes in which the
flux is F$_{\rm 0.3-10\,keV} \sim$15$\times$10$^{-13}$ erg cm$^{-2}$ s$^{-1}$,
interleaved by low-activity state. During the X-ray high-activity state in 2013
November, the
$\gamma$-ray flux is roughly 2.6 times the average value, whereas in 2013
August--October the source is not detected in $\gamma$-rays. No
significant X-ray photon index variability is observed
(Fig. \ref{MWL}).\\
In the five months between
the {\it XMM-Newton} observations the source brightened by about 35 per cent.
The photon index derived for the two {\it XMM-Newton} observations is in
good agreement with the {\it Swift}-XRT results, and is
slightly softer than the value obtained in 2001 by \citet{gallo06}. This may
be due to the higher flux observed in 2001 (F$_{0.3-10\,keV} =
1.5 \times 10^{-12}$ erg cm$^{-2}$ s$^{-1}$) than that observed in
the 2012 {\it XMM-Newton} 
observations. In the 2012 {\it XMM-Newton} observations, there is no
evidence of soft X-ray excess.\\
%
%
%

\subsubsection{Optical and UV properties}

On monthly time-scales a difference of $\sim$0.3 mag was observed for
PKS\,2004$-$447 in $u$ band 
by {\em XMM}-OM. During the {\em Swift}-UVOT observations the change of
magnitudes spanned in the various band is about 0.6, 1.3, 1.0, 0.9, 0.7, 0.8 
going from the $v$ to $w2$ filter (Table \ref{uvot}). A first peak of the
optical/UV activity was observed on 2011 September 5, during a high X-ray and
$\gamma$-ray activity period. A second peak of activity was observed in the $u$ and
$w1$ bands on 2013 November 19 and 20. At that time, the
maximum X-ray flux was observed together with a high $\gamma$-ray flux
level. A contemporaneous increase of activity between the optical-UV and the
X-ray and $\gamma$-ray bands indicates that the jet emission is dominant also in
optical and UV, in agreement with the lack of a significant disc
emission suggested 
by \citet{abdo09}.

\section{Discussion}

\subsection{Relativistic jets in RL-NLSy1}

High-energy emission has been detected in RL-NLSy1.
On the other hand, no $\gamma$-ray emission has been found in RQ-NLSy1,
suggesting a possible intrinsically different nature between these two
sub-populations. Different $\gamma$-ray properties have been observed
in the six RL-NLSy1 detected by {\it Fermi}-LAT.
Three objects,
PMN\,J0948$+$0022, SBS\,0846$+$513, and 1H\,0323$+$342 show
$\gamma$-ray flares, reaching an intrinsic apparent luminosity as
high as 10$^{48}$ erg s$^{-1}$ 
\citep{dammando15,dammando12}, i.e. comparable to those shown by FSRQ
\citep[e.g.][]{ackermann11}.
On the other hand, PKS\,2004$-$447, PKS\,1502$+$036, and
FBQS\,J1644$+$2619 have not shown strong flares so far.\\
During the first 6 years of
{\it Fermi} operation,
%
%
the apparent luminosity of PKS\,2004$-$447 ranges between $L_{\gamma}
= ($1.3--4.2)$\times$10$^{45}$ 
erg s$^{-1}$. A similar 
behaviour was found for 
PKS\,1502$+$036 with a slightly larger luminosity
\citep[$L_{\gamma} \sim 10^{46}$
erg s$^{-1}$,][]{dammando13}. FBQS\,J1644+2619 had an
intermediate behaviour with high activity episodes interleaved by long
quiescent periods \citep{dammando15b}.\\
The RL-NLSy1 J1548$+$3511 is not detected in $\gamma$-rays and the upper
limit to the apparent isotropic luminosity is $L_{\gamma} < 1.7 \times
10^{46}$ erg s$^{-1}$. \\
The detection of $\gamma$-ray emission in a handful of RL-NLSy1 proves
the presence of relativistic jets in this peculiar sub-class of active
galactic nuclei (AGN). In the
photon index versus $\gamma$-ray luminosity plane, the $\gamma$-ray
loud NLSy1 are 
located in the low-luminosity tail of FSRQ distribution, where
also the $\gamma$-ray emitting steep-spectrum radio quasars are found
\citep{ackermann15, abdo10}. \\
A jet component contribution is likely observed above 2 keV in the
RL-NLSy1 J1548$+$3511, where the X-ray photon index is $\Gamma_{\rm X} \sim
1.7$--1.9. Below 2 keV the spectrum is 
softer, compatible with the presence of soft X-ray excess usually
observed in 
NLSy1 \citep{grupe10}, as well as in the $\gamma$-ray NLSy1
PMN\,J0948$+$0022 \citep{dammando14} and 1H\,0323$+$342
\citep{paliya14}. An indication of a weak soft excess was reported by
\citet{gallo06} for PKS\,2004$-$447 in 2004.
The X-ray spectra of
PKS\,2004$-$447 observed in 2012 are well reproduced by a single power
law with hard 
photon index $\Gamma_{\rm X} \sim 1.7$ and no significant soft X-ray
excess is needed below 2 keV. 
%
%
This may be due to the relatively low
statistics of the 2012 observations.
%
%
%
The study of the multiwavelength variability of PKS\,2004$-$447
indicates that during the high activity state observed in
$\gamma$-rays between 2011 August and October, when the $\gamma$-ray
flux was a factor of 2.6 times the average value, also the X-ray, UV and optical
emission reached a maximum, suggesting a common origin for the
multiband variability. It is worth mentioning that not all the episodes
of flux increase in X-rays/UV
are associated with a high activity state in $\gamma$-rays
(Fig. \ref{MWL}), like in the case of the high X-ray activity observed
in 2013 September. Such X-ray/UV flares with no obvious counterpart in
other bands were observed in the RL-NLSy1
PMN\,J0948$+$0022 \citep[e.g.][]{foschini12}, as well as in many blazars
\citep[e.g., 3C\,279,][]{abdo10b}.\\
No systematic studies of complete
samples of NLSy1 have been carried out at Very High Energy
(VHE) so far.
PKS\,2004$-$447 was observed by the High Energy Stereoscopic System
\citep[H.E.S.S.,][]{aharonian06}. No detection was obtained and the
estimated upper limit was 0.9 per cent of the Crab Units
\citep{abramowski14}. Upper limits of 10 per cent and 1.9
per cent of the Crab Units at VHE were obtained for the NLSy1
1H\,0323$+$342 \citep{falcone04} and PMN\,J0948$+$0022
\citep{dammando15}, respectively. The lack of VHE detection may be due
either to an intrinsically soft $\gamma$-ray spectrum, or to
$\gamma$-$\gamma$ absorption within the source, or to pair
production from 
$\gamma$-ray photons of the source and the infrared photons from the
extragalactic background light (EBL), although the latter scenario is
disfavoured by the fact that
the redshift of the
most distant FSRQ detected at VHE, PKS\,1441$+$25 \citep{mirzoyan15,
  mukherjee15}, is higher ($z=0.939$) than the redshifts of the RL-NLSy1
investigated so far.\\

\subsection{Physical properties}

Milliarcsecond resolution observations are a fundamental requirement
for describing 
the morphology and understanding the physical properties of
RL-NLSy1. Given their compactness on arcsecond scale, the high angular
resolution provided by VLBA observations allows us to investigate the
presence of a jet structure emerging from the core region and to
constrain the physical characteristics of the core emission without
substantial contamination from the jet.\\
We computed the brightness temperature $T_{\rm B}$ of the core
component by using:

\begin{equation}
T_{\rm B} = \frac{1}{2 k} \frac{S(\nu)}{\Omega} \left( \frac{c}{\nu} \right)^{2}
\label{eq_brightness}
\end{equation}

\noindent where $k$ is the Boltzmann constant, $\Omega$ is the solid
angle of the emitting regions, $S(\nu)$ is the source-frame flux density at the
observed frequency, and $c$ is the speed of light. We computed the
source-frame flux density as
$S(\nu) = S_{\rm obs}(\nu) \times (1+z)^{1-\alpha}$, where $S_{\rm
  obs}(\nu)$ is the observer-frame flux density, $z$ is the source
redshift, and $\alpha$ is the radio spectral index that we assume to be
equal to 0 for the self-absorbed core component. The solid angle is
given by:

\begin{equation}
\Omega = \frac{\pi}{4} \theta_{\rm min} \theta_{\rm maj}
\label{omega}
\end{equation}

\noindent where $\theta_{\rm maj}$ and $\theta_{\rm min}$ are the major axis
and minor axis, respectively. In both sources the core region is
unresolved by our VLBA observations and the major and minor axes
must be considered as upper 
limits. This ends up in a lower limit to the brightness
temperature.\\
If in Eq. \ref{eq_brightness} we
consider the values derived from the VLBA image at 15 GHz, we obtain a
brightness temperature $T_{\rm B} > 4.4\times 10^{9}$ K for
J1548$+$3511. In the case of PKS\,2004$-$447 we computed the core
brightness temperature making use of the values derived from the VLBA
image at 1.4 GHz, and we obtained $T_{\rm B} > 2\times 10^{10}$ K.\\ 

For PKS\,2004$-$447, the availability of two VLA observations
separated by about one month allowed us to estimate the rest-frame
variability brightness temperature, $T'_{\rm B, var}$ by using the flux density
variability. Following \citet{dammando13} we computed $T'_{\rm B}$ by:

\begin{equation}
T'_{\rm B, var} = \frac{2}{\pi k} \frac{|\Delta S| D_{\rm L}^{2}}{\Delta
  t^{2} \nu^{2} (1+z)^{1 + \alpha}}
\label{var_bright}
\end{equation}

\noindent where $|\Delta S|$ is the flux density variation, $\Delta t$
is the time lag between the observations, and $D_{\rm L}$ is the
luminosity distance. If in Eq. \ref{var_bright} we consider $|\Delta
S|=221$ mJy and $\Delta t = 36$ days, i.e. the flux density variation
measured between the last two VLA observations, we obtain $T'_{\rm B, var}
\sim 1.7 \times 10^{14}$K. This high variability brightness
temperature is similar 
to the value derived by \citet{gallo06} on the basis of Ceduna
monitoring campaign. In the case of J1548$+$3511, \citet{yuan08}
estimated a variability brightness temperature $T'_{\rm B, var} =
10^{13}$K on the basis of the flux density variation measured in 207
days at 4.9 GHz. 
Assuming that such high values are due to Doppler
boosting, we can estimate the variability Doppler factor $\delta_{\rm
  var}$, by using:

\begin{equation}
\delta_{\rm var} = \left( \frac{T'_{\rm B, var}}{T_{\rm int}} \right)^{\frac{1}{3+\alpha}}
\label{doppler}
\end{equation}

\noindent where $T_{\rm int}$ is the intrinsic brightness
temperature. Assuming a typical value $T_{\rm int}=5 \times 10^{10}$K,
as derived by e.g. \citet{readhead94} and \citet{hovatta09}, and a flat spectrum
with $\alpha=0$, we obtain $\delta =15$ and 5.8 for PKS\,2004$-$447
and J1548$+$3511, respectively. These 
values are in agreement with the range of variability Doppler 
factor found for the RL-NLSy1 SBS\,0846$+$513 \citep{dammando13b},
PKS\,1502$+$036 \citep{dammando13}, as well as for blazars
\citep{savolainen10}.\\ 
We estimated the ranges of viewing angles $\theta$ and of the bulk
velocity in terms of speed of light 
$\beta$ from the
jet/counter-jet brightness ratio. Assuming that the source has two
symmetrical jets of the same intrinsic power, we used the equation:

\begin{equation}
\frac{B_{\rm j}}{B_{\rm cj}} = \left( \frac{1 + \beta {\rm
      cos}\theta}{1 - \beta {\rm cos}\theta} \right)^{2 + \alpha}
\label{brightness}
\end{equation}

\noindent where $B_{\rm j}$ and $B_{\rm cj}$ are the jet and counter-jet
brightness, respectively. We prefer to compare the surface brightness
instead of the flux density because the jet has a smooth structure
without clear knots. The jet brightness for J1548$+$3511 and
PKS\,2004$-$447 is 0.93 mJy beam$^{-1}$ and 14.0 mJy beam$^{-1}$, respectively.
In the case of the counter-jet, which is not visible, we assumed an
upper limit for the surface brightness that corresponds to 0.2
mJy beam$^{-1}$ and 0.6 mJy beam$^{-1}$ for J1548$+$3511 and PKS\,2004$-$447,
respectively, i.e. 1$\sigma$ noise level measured on the image. 
From the brightness ratio estimated from Eq. \ref{brightness} we
obtain $\beta$cos$\theta > 0.61$ for J1548$+$3511 and 0.52 for
PKS\,2004$-$447, implying that the minimum velocity is $\beta > 0.6$
and 0.5 
%
%
and a
maximum viewing angle $\theta = 52^{\circ}$ and 59$^{\circ}$, respectively. 
These limits do not provide a tight constraint on the physical
parameters of these objects.\\


\subsection{SED modelling}

\begin{figure*}
\begin{center}
\includegraphics{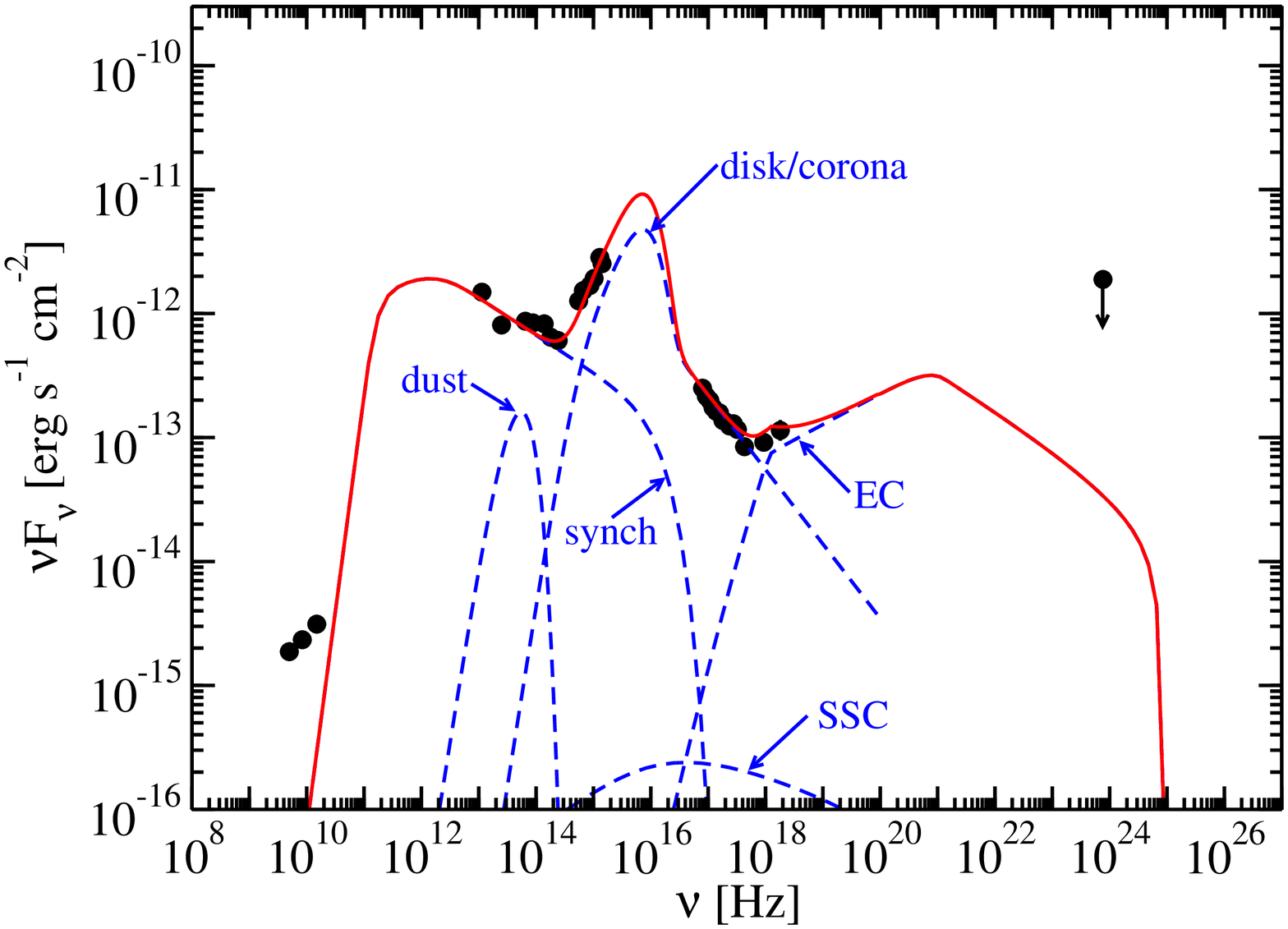}
\includegraphics{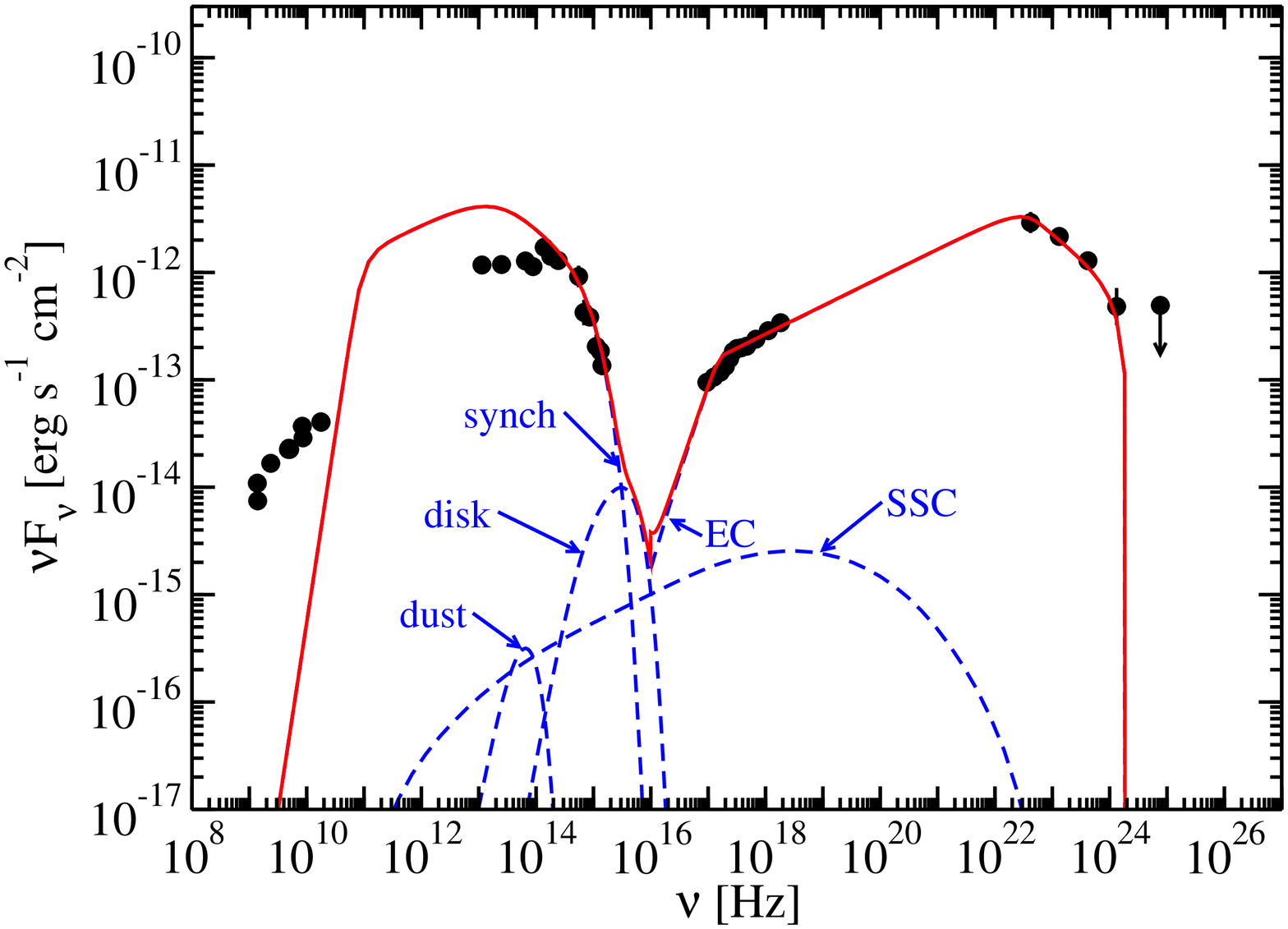}
\vspace{8.5cm}
\caption{SED data (circles) and model fit (solid curve) of
  J1548$+$3511 ({\it left}) and PKS\,2004$-$447 ({\it right}). The
  model components are shown as dashed curves. Multiwavelength data
  are not simultaneous.}
\label{SED_fig}
\end{center}
\end{figure*}

We created an average SED for the two NLSy1 studied here. The SED of
PKS\,2004$-$447 includes the 6-year average {\em Fermi}-LAT spectrum,
the {\it
XMM-Newton} EPIC-pn data collected on 2012 October 18, and the {\it
Swift}-UVOT data collected on 2012 September 12. In addition, we
included the IR data collected by WISE on 2010 April 14 and by Siding
Spring Observatory in 2004 April 10 -- 12, the radio VLA and
VLBA data presented in this paper, and the ATCA data
from \citet{gallo06}. 
The SED of J1548$+$3511 includes the upper limit estimated
over 6 years of {\it Fermi} observations, the {\it XMM-Newton} (EPIC-pn and
OM) data collected on 2011 August 8, the IR data collected by WISE on
2010 January 30 and by 2MASS 
on 1998 April 3, and the radio VLA data presented here.
The multiwavelength data are not simultaneous.\\
We modelled the two SED with a combination of synchrotron, synchrotron
self-Compton (SSC), and external Compton (EC) non-thermal emission. The
synchrotron component considered is self-absorbed below $10^{11}$\ Hz and
thus cannot reproduce the radio emission. This emission is likely from the
superposition of multiple self-absorbed jet components \citep{konigl81}. 
We also included thermal emission by an accretion disc and dust torus. 
The modelling details can be found in \citet{finke08} and
\citet{dermer09}. Additionally, a soft excess was observed in the X-ray
spectrum of J1548$+$3511. We note that the origin of this soft X-ray
emission is uncertain.
In order to account for this feature, as a possible origin, we
included emission from the disc, which is Compton scattered by an optically
thin thermal plasma near the accretion disc (i.e., a corona). This was
done using the routine ``{\tt SIMPL}'' \citep{steiner09}.  This routine
has two free parameters: the fraction of disc photons scattered by the
corona ($f_{\rm sc}$), and the power-law photon index of the scattered coronal
emission ($\Gamma_{\rm sc}$). The mass of the BH was chosen to be the same as
the one reported in \citet{foschini15}.\\
The results of the modelling are reported in Table \ref{table_fit} and
Figure \ref{SED_fig} \citep[for a description of the model parameters
see][]{dermer09}. This model assumes that the emitting region is outside the
Broad Line Region, where dust torus photons are likely the seed
photon source. 
This seed photon source was modelled as being an isotropic,
monochromatic radiation 
source with dust parameters chosen to be consistent with the relation
between inner radius, disc luminosity, and dust temperature from
\citet{nenkova08}. \\
The Compton component of the SED of PKS\,2004$-$447 and J1548$+$3511
is modelled with an external Compton scattering of dust torus seed
photons, as for the SED of SBS\,0846$+$513 \citep{dammando13} and
PMN\,J0948$+$0022 \citep{dammando15}. 
In PKS\,2004$-$447 the IR data are not well fitted by the
model. This may be due to the fact that the data considered in the SED
are not simultaneous and some variability may be present (note that
these data are not well-fitted by the modelling done by
\citet{paliya13} either). We note that
the disc
luminosity of PKS\,2004$-$447 is particularly weak. 
This value is similar to the disc luminosity of SBS\,0846$+$513, while
it is significantly lower than that of PMN\,J0948$+$0022.
The disc luminosity of PKS\,2004$-$447 is consistent
with the value estimated by \citet{foschini15} on the basis of the
optical spectrum, and the lack of a blue
bump. This is in contrast to the modelling of PKS\,2004$-$447 by
\citet{paliya13} who used a much brighter dust component, which  
was not consistent with the disc luminosity estimated by \citet{foschini15}.
It is worth mentioning that the model parameters shown here are not
unique, and other model parameters could reproduce the SED equally well. This is
particularly true for J1548$+$3511, 
since without a LAT detection, its Compton component is not
well-constrained.\\

\begin{table*}
\footnotesize
\begin{center}
\caption{Parameters used to model the SED.}
\label{table_fit}
\begin{tabular}{lcccc}
\hline
Parameter & Symbol & PKS\,2004$-$447  & J1548$+$3511 \\
\hline
Redshift & 	$z$	& 0.24 & 0.478	\\
Bulk Lorentz Factor & $\Gamma$	& 30 & 30 \\
Doppler factor & $\delta_D$	& 30 & 30 \\
Magnetic Field [G]& $B$         & 0.5 & 2.0 \\
Variability Timescale [s]& $t_v$       & $10^5$ & $10^5$ \\
Comoving radius of blob [cm]& $R^{\prime}_b$ & 7.3$\times$10$^{16}$ & $6.1\times10^{16}$ \\
\hline
Low-Energy Electron Spectral Index & $p_1$       & 2.5 & 2.5 \\
High-Energy Electron Spectral Index  & $p_2$       & 3.8 & 3.6\\
Minimum Electron Lorentz Factor & $\gamma^{\prime}_{\rm min}$  & 1.0 & 3.0 \\
Break Electron Lorentz Factor & $\gamma^{\prime}_{\rm brk}$ & $6.0\times10^2$ & $1.0\times10^2$\\
Maximum Electron Lorentz Factor & $\gamma^{\prime}_{\rm max}$  & $4.0\times10^3$ & $1.0\times10^4$ \\
\hline
Black hole mass [$M_\odot$] & $M_{\rm BH}$ & $4.3\times10^6$ & $8.3\times10^7$ \\
Disc luminosity [$\erg\ \s^{-1}$] & $L_{\rm disc}$ & $1.8\times10^{42}$ & $1.4\times10^{45}$ \\
Inner disc radius [$R_g$] & $R_{\rm in}$ &  6.0 & 2.0\\
Outer disc radius [$R_g$] & $R_{\rm out}$ &  200 & 200 \\
\hline
Fraction of disc photons scattered by corona & $f_{\rm sc}$ & 0 & 0.08 \\
Corona spectral index & $\G_{\rm sc}$ & N/A & 2.6 \\
\hline
Seed photon source energy density [$\erg\ \cm^{-3}$] & $u_{\rm seed}$ & $8.3\times10^{-6}$ & $2.7\times10^{-5}$ \\
Seed photon source photon energy & $\e_{\rm seed}$ & $5\times10^{-7}$ & $5\times10^{-7}$ \\
Dust torus luminosity [$\erg\ \s^{-1}$] & $L_{\rm dust}$ & $7.5\times10^{40}$ & $1.9\times10^{44}$ \\
Dust torus radius [cm] & $R_{\rm dust}$ & $6.5\times10^{15}$ & $4.1\times10^{18}$ \\
Dust temperature [K] & $T_{\rm dust}$ & 1000 & 1000 \\
\hline
Jet Power in Magnetic Field [$\erg\ \s^{-1}$] & $P_{\rm j,B}$ & $8.9\times10^{45}$ & $1.0\times10^{47}$\\
Jet Power in Electrons [$\erg\ \s^{-1}$] & $P_{\rm j,e}$ & $2.7\times10^{44}$ & $6.6\times10^{43}$ \\
\hline
\end{tabular}
\end{center}
\end{table*}

\subsection{RL-NLSy1 and the young radio source population}

The low black hole mass and the high accretion rate commonly estimated in
NLSy1 suggested that these objects may be in an early evolutionary
stage \citep{grupe99, mathur00}. 
Some RL-NLSy1 have been proposed to be compact steep
spectrum (CSS) radio 
sources, on the basis of their compactness and flat/inverted spectrum
which turns over at a few hundred MHz or around one GHz \citep{yuan08}. CSS are
powerful radio sources whose linear size is $<20$ kpc. Due to their
intrinsically compact size they are considered radio sources in an
early evolutionary stage (see e.g. O'Dea 1998 for a review). Kinematic
and radiative studies provided ages of 10$^{3-5}$ years, strongly
favouring the youth scenario \citep[see e.g.][]{polatidis03,murgia03}. \\
Many CSS sources are hosted in galaxies that recently underwent major mergers,
which may have triggered the onset of the radio emission. A similar
scenario was suggested by \citet{leon14} for the RL-NLSy1
1H\,0323$+$342. A link between NLSy1 and CSS was also proposed by
\citet{caccianiga14} for the RL-NLSy1 SDSS\,J143244.91$+$301435.3 on the
basis of its compact size, absence of variability, and a steep radio spectrum. 
In this context the variable and $\gamma$-ray emitting RL-NLSy1 may be
the aligned population of RL-NLSy1 where the CSS properties are hidden
by dominant boosting effects. However, at least
three (1H\,0323$+$342, PMN\,J0948$+$0022, and FBQS\,J1644$+$2619) out of the six RL-NLSy1 detected in $\gamma$-rays have extended
structures with linear size of 20-50 kpc \citep{doi12}, challenging
this interpretation. Among the remaining three objects,
PKS\,2004$-$447 was suggested as a possible CSS source by
\citet{gallo06}. However, the X-ray spectra in CSS sources are
typically highly obscured with column density N$_{\rm H} \gtrsim
10^{22}$ cm$^{-2}$ \citep{tengstrand09}, while no absorber in addition
to the Galactic one is needed for modelling the X-ray spectrum of
PKS\,2004$-$447.\\ 
The information available so far is not enough to firmly link the
RL-NLSy1 to the young radio source population. Statistical
multiwavelength studies on a large sample of RL-NLSy1 are required for
investigating a possible connection between these sub-class of radio-loud AGN.\\
 
\section{Conclusions}

We presented results on a multiwavelength study, from radio to
$\gamma$-rays, of the RL-NLSy1 J1548$+$3511 and PKS\,2004$-$447. The
conclusions from this investigation can be summarized as follows:

\begin{itemize}

\item PKS\,2004$-$447 is detected in $\gamma$-rays by LAT with an
  average flux between 0.1 and 100 GeV of 
  $\sim$1.6$\times$10$^{-8}$ ph\,cm$^{-2}$ s$^{-1}$, corresponding to a
  luminosity of $\sim$1.6$\times$10$^{45}$ erg s$^{-1}$, which is comparable
  to the values found in the other $\gamma$-ray emitting NLSy1. No
  strong flares have been detected so far. \\

\item J1548$+$3511 has not been detected in $\gamma$-rays by LAT during the
  first 6 years of observations. The upper limit to the
  luminosity is 1.7$\times$10$^{46}$ erg/s.\\

\item Both sources have a clear core-jet structure on parsec
  scales. The majority of the radio emission comes from the core
  component. On arcsecond scale the radio structure is
  unresolved. \\

\item PKS\,2004$-$447 shows significant variability from
  radio to $\gamma$-rays. In particular, at the end of 2011 and at the
  end of 2013 the high activity state observed in $\gamma$-rays is
  simultaneous to a local maximum in X-rays, UV and optical
  bands, suggesting a common origin. The X-ray spectra collected by
  {\it XMM-Newton} in 2012 from 0.3 to 10
  keV are fitted by a single power law. No significant X-ray
  photon index variability is observed in the period considered in
  this paper.\\

\item The X-ray spectrum of J1548$+$3511 is well fitted by a soft
  component at low energy, and by a hard component above 2 keV. The X-ray flux
  increases of about 25 per cent in the 
  two weeks between the {\it XMM-Newton} observations. The brightening
  is accompanied by a hardening of the spectrum, in particular above
  the energy break. This is a further indication that the emission in
  the high-energy part of the X-ray spectrum is dominated by the
  non-thermal jet emission, while a Seyfert component may be present in
  the low-energy part of the spectrum.\\

\item The broadband SED of both sources can be
reproduced with synchrotron, SSC and external Compton scattering of
the IR seed photons from the dust torus. The soft excess in
J1548$+$3511 can be 
modelled as emission from a thermal corona. The disc luminosity of
PKS\,2004$-$447 turns out to be very weak.\\

\item The variability brightness temperature and the Doppler factor
  derived for both sources are similar to those found in $\gamma$-ray
  blazars. These characteristics, together with the high
  radio-loudness are a good proxy for the presence of relativistic
  jet. However, they are not good tools in the selection of
  $\gamma$-ray emitting NLSy1.\\

\end{itemize}

\section*{Acknowledgments}

We thank the referee Dirk Grupe for helpful and valuable comments that improved the manuscript. 
The VLBA and VLA are operated by the US National Radio Astronomy Observatory
which is a facility of the National Science Foundation operated under
a cooperative agreement by Associated Universities, Inc. \\ 
The {\it Fermi} LAT Collaboration acknowledges generous ongoing support
from a number of agencies and institutes that have supported both the
development and the operation of the LAT as well as scientific data analysis.
These include the National Aeronautics and Space Administration and the
Department of Energy in the United States, the Commissariat \`a
l'Energie Atomique 
and the Centre National de la Recherche Scientifique / Institut
National de Physique 
Nucl\'eaire et de Physique des Particules in France, the Agenzia
Spaziale Italiana 
and the Istituto Nazionale di Fisica Nucleare in Italy, the Ministry
of Education, 
Culture, Sports, Science and Technology (MEXT), High Energy Accelerator Research
Organization (KEK) and Japan Aerospace Exploration Agency (JAXA) in Japan, and
the K.~A.~Wallenberg Foundation, the Swedish Research Council and the
Swedish National Space Board in Sweden. Additional support for science analysis during the operations phase is gratefully
acknowledged from the Istituto Nazionale di Astrofisica in Italy and the Centre National d'\'Etudes Spatiales in France.\\
This research has made use of the NASA/IPAC
Extragalactic Database NED which is operated by the JPL, California
Institute of Technology, under contract with the National Aeronautics
and Space Administration.
This publication makes use of data products from the Wide-field Infrared
Survey Explorer, which is a joint project of the University of California,
Los Angeles, and the Jet Propulsion Laboratory/California Institute of
Technology, and NEOWISE, which is a project of the Jet Propulsion
Laboratory/California Institute of Technology. WISE and NEOWISE are funded
by the National Aeronautics and Space Administration.
This publication makes use of data products from the Two Micron All Sky
Survey, which is a joint project of the University of Massachusetts and
the Infrared Processing and Analysis Center/California Institute of
Technology, funded by the National Aeronautics and Space Administration
and the National Science Foundation.

\end{document}